\RequirePackage{ifpdf}
\RequirePackage{fixltx2e}
\documentclass[a4paper,12pt]{JHEP3}
\pdfoutput=1
\usepackage[latin1]{inputenc}

\usepackage{amsmath}
\usepackage{amsfonts}
\usepackage{amssymb}
\usepackage{graphicx}

\usepackage{mathrsfs}
\usepackage{listings}
\usepackage{verbatim}
\usepackage{mathtools}
\usepackage{simplewick}
\usepackage{color}\let\normalcolor\relax
\usepackage{url}

\newcommand{\bomega}{\boldsymbol{\omega}}

\usepackage{subfig}


\newcommand{\be}{\begin{equation}}
\newcommand{\ee}{\end{equation}}
\newcommand{\bq}{\begin{eqnarray}}
\newcommand{\eq}{\end{eqnarray}}

\newcommand{\one}{\hbox{\rm 1\kern-.27em I}}

\newcommand{\pdfannotlink}{\pdfstartlink}

\title{Contact interactions between particle worldlines}
\author{James P. Edwards \\Centre for Particle
Theory, Department of Mathematical Sciences, \\
University of Durham, Durham DH1 3LE, UK \\ Email:
j.p.edwards@durham.ac.uk}

\abstract{We construct contact interactions for bosonic and fermionic point particles. We first relate the resulting theories to classical electrostatics by taking functional averages over worldlines whose endpoints are fixed to charged particles. Counting those paths which pass through a space-time point $x^{\mu}$ gives the static electric field at that point, provided we take the limit where the length measured along the worldlines is large. We also investigate corrections to the classical field that arise beyond leading order in this limit before constructing a theory of point particles that interact when their worldlines intersect. We quantise this theory and show that the partition function contains propagator couplings between the endpoints of the particles before discussing how this is related to the worldline formalism of quantum field theory and general action at a distance theories.}

\keywords{Effective field theories, Field theories in lower dimensions, Confinement}

\preprint{DCPT-15/39}

\begin{document}

\abstract{We construct contact interactions for bosonic and spinning strings. In the tensionless limit of the spinning string this reproduces the super-Wilson loop that couples spinor matter to Abelian gauge theory. Adding boundary terms that quantise the motion of charges results in a string model  equivalent to spinor QED. The strings represent lines of electric flux connected to the charges. The purely bosonic model is spoilt by divergences that are excluded from the spinning model by world-sheet supersymmetry.}

\section{Introduction}
Classical electromagnetism is conventionally described by Maxwell's field theory and there seems to be little room for debate about its formulation. In \cite{Us1} and \cite{Us2}, however, building upon \cite{Paul1} an alternative approach to determining the electromagnetic field strength tensor for a pair of charged particles led directly to a novel interacting string theory. This theory contained contact interactions on the string worldsheet which served to produce expectation values of Wilson lines in Abelian quantum field theory. In the case of electrostatics, the description given in \cite{Paul1} was in terms of point particles whose worldlines have their endpoints fixed to the charged particles. The electric field at a position in space-time was arrived at via a weighted average over all such worldlines which also pass through the given point. The physical picture which motivated this approach (and the later interacting string theory) is of Faraday's lines of force as fundamental objects which become the physical degrees of freedom of the electromagnetic field.

To complement this work on contact interactions in string theory it seems appropriate to return to worldline theories to explore the consequences of allowing point particles to interact when their worldlines intersect. Such theories are of significant physical interest, since the so-called worldline formalism of quantum field theory \cite{Strass1, Strass2, WL3} expresses physical quantities in a field theory in terms of the quantum mechanics of point particles which trace out worldlines in space-time \cite{Schu, Auer, Anton}. This technique can also be extended to non-Abelian theories and chiral fermions \cite{Paul2} where it may provide insight to the unification of the fundamental forces \cite{Me}. The coupling between matter fields and the gauge field is described in the point particle theory as a Wilson-line interaction for the particle worldlines in the presence of a background field. For example, the partition function for a single scalar field minimally coupled to an Abelian gauge field, $\mathcal{A}$, is expressed in the worldline formalism (which we derive in more detail later in the paper) by an integral over all closed curves $\omega$
\begin{equation}
	\mathcal{Z} = \int_{0}^{\infty}\frac{dT}{T} \oint \mathscr{D}\omega \,e^{-S_{\textrm{point}} \left[\omega\right]} W\left[\mathcal{A}\right]\,; \qquad W\left[\mathcal{A}\right] = e^{i \int d\omega \cdot \mathcal{A}},
\end{equation}
where $S_{\textrm{point}}\left[\omega\right] = m\int d\tau \sqrt{\dot{\omega}^{2}}$ is an action describing the dynamics of a point particle and $W\left[\mathcal{A}\right]$ is the Wilson loop describing the interaction of the particle with the gauge field (we have absorbed the coupling strength into $\mathcal{A}$). The right hand side is interpreted as quantum mechanics on the worldline of this particle and it is this first quantised theory which we propose to modify in this paper. 

Field theory is the conventional framework in which to introduce interactions and the local nature of this approach naturally leads to particles interacting upon contact. However, the worldline formalism can offer substantial calculational advantages over traditional approaches in field theory, especially since it represents a reorganisation of the usual perturbative expansion in Feynman diagrams and makes the local gauge invariance manifest \cite{Diet}. It is therefore important to develop worldline techniques further and one of the most basic modifications to the theory must be to introduce direct interactions between these particle worldlines. As we shall describe below, a modification to the worldline theory can be interpreted as inducing a change in the underlying field theory, so the results of this program may provide new tools to complement the conventional techniques familiar to field theorists. We comment on this in section \ref{contact}.

Direct inter-particle interactions can be found in many previous publications. One of the most well-known approaches is the action at a distance formulation of electrodynamics by Feynman and Wheeler \cite{FeynWheel, Kerner}, originally proposed to address the problem of radiation reaction. This built upon earlier work in formulating a consistent theory involving direct inter-particle interactions by Tetrode \cite{Tetrode} and Fokker \cite{Fokker}, who described electromagnetic phenomena in terms of interactions between particles with light-like separation. Ramond generalised this work and found a set of consistency constraints limiting the form of the interaction that can be introduced into worldline theories \cite{KalbR2}. This was further extended to include direct inter-string interactions as well as interactions between particles and strings \cite{KalbR, Letel1, Letel2, NRH, Baker}. A number of further theories involving action at a distance have been proposed to describe various other phenomena within this framework -- see for example \cite{Katz1, Katz2, Onem, Wigner, Dett}.

The general principle is to couple the worldlines of the particles together by adding extra terms to the free particle action, $S_{\textrm{point}}$. An illustrative example would be to consider a theory of two particles whose worldlines are described by $\omega_{a}^{\mu}\left(\tau_{a}\right)$ and $\omega_{b}^{\mu}\left(\tau_{b}\right)$ and to introduce \cite{mech, Onem}
\begin{equation}
	S_{\textrm{int}} = g_{a}g_{b} \int_{\omega_{a}} d\tau_{a} \int_{\omega_{b}} d\tau_{b}~ \dot{\omega}_{a}\left(\tau_{a}\right) \cdot \dot{\omega}_{b}\left(\tau_{b}\right) D\left(\omega_{a} - \omega_{b}\right)
	\label{Gint}
\end{equation}
as an interaction term in the action\footnote{To be precise this provides vector-like interactions between point particles. Scalar interactions can be produced by replacing each $\dot{\omega_{i}}^{\mu}$ by $\left(\dot{\omega}_{i}^{2}\right)^{\frac{1}{2}}$.}. Here the function $D$ must be symmetric and accounts for the relative strength of interaction as a function of particle separation. In the literature discussed above $D\left(\omega_{a} - \omega_{b}\right)$ has been taken to be supported for light-like, time-like and space-like separations, the last of which is the relativistic generalisation of an instantaneous interaction. In Feynman-Wheeler theory, for example, $D\left(\omega_{a} - \omega_{b}\right)$ is taken to be the sum of advanced and retarded Green functions of the (space-time) Laplacian. As is now well known, these action at a distance theories are often cast into a form reminiscent of a field theory, although the ``fields'' are not independent variables but are rather defined in terms of the particle trajectories and the choice of $D$. For instance, if we take \cite{Barut}
\begin{equation}
	D\left(\omega_{a} - \omega_{b}\right) = \frac{1}{4\pi}\delta\left(\left(\omega_{a} - \omega_{b}\right)^{2}\right) - \frac{m}{8\pi}\frac{\theta\left(\left(\omega_{a} - \omega_{b}\right)^{2}\right)}{\sqrt{\left(\omega_{a} - \omega_{b}\right)^{2}}}J_{1}\left(m \sqrt{\left(\omega_{a} - \omega_{b}\right)^{2}}\right)
\end{equation}
which is the time symmetric Green satisfying $\left(\partial_{\mu}\partial^{\mu} - m^{2}\right)D\left(\omega_{a} - \omega_{b}\right) = -\delta^{4}\left(\omega_{a} - \omega_{b}\right)$ then we may define
\begin{equation}	
	A^{\mu}\left(x\right) =  \sum_{a}g_{a} \int d\tau_{a}\, \dot{\omega}^{\mu}_{a}\left(\tau_{a}\right) D\left(\omega_{a}\left(\tau_{a}\right) - x\right).
\end{equation} 
This satisfies the Maxwell equations and gauge condition
\begin{align}
	\left(\partial_{\nu}\partial^{\nu} - m^{2}\right)A^{\mu}\left(x\right) &= j^{\mu}\left(x\right); \qquad j^{\mu}\left(x\right) = -\sum_{a}g_{a} \int d\tau_{a} \, \dot{\omega}^{\mu}_{a}\left(\tau_{a}\right) \delta^{4}\left(x - \omega_{a}\right)\,,\nonumber \\
	\partial_{\mu}A^{\mu}\left(x\right) &= 0
\end{align}
and in terms of $A$ the interaction between the particles takes the form
\begin{equation}
	\sum_{a} g_{a} \int_{\omega_{a}} d\tau_{a}\, \dot{\omega}_{a}\left(\tau_{a}\right) \cdot A\left(\omega_{a}\left(\tau_{a}\right)\right)
\end{equation}
which shows that the action at a distance formalism contains the same equations of motion and interactions as more traditional approaches using field theory. 

The proposal we will make will follow the same form as (\ref{Gint}) except that we shall choose $D\left(\omega_{a} - \omega_{b}\right) = \delta^{4}\left(\omega_{a} - \omega_{b}\right)$ so as to provide contact interactions between the worldlines. This also ensures that, although in principle (\ref{Gint}) implies the interaction is non-local on the worldlines, the particles only communicate when they meet so that the theory is local in space-time. In other words we are no longer considering action at a distance but we allow particles to interact when they find themselves at the same space-time position. As stated above, we have previously considered such contact interactions between strings, where the theory found application to classical electromagnetism and quantum electrodynamics. We now intend to explore the same ideas for the case of point particles.

This article revisits and extends the results of \cite{Paul1} and also generalises that work to the case of fermionic particles. It then goes beyond leading order in the coupling strength to demonstrate that in fact the full quantum theory of a set of interacting point particles is consistent and free of unwanted divergences. We develop the functional approach to one dimensional field theory for consistency with \cite{Us1, Us2} and for the generalisation to fermionic particles we will find it most natural to form the theory in superspace. We will first consider a single particle worldline with fixed endpoints that is constrained to pass through a given point in space and will relate it to classical electrostatics and the well-known phenomenon of confinement. We then repeat a similar calculation for spin $1/2$ particles to explore the fermionic version of the theory before considering an arbitrary set of interacting worldlines. We will see that the partition function of this theory is related to propagators of the Klein-Gordon operator.

The structure of this article is as follows. In section \ref{Bosonic} the bosonic theory presented at lowest order in \cite{Paul1} is reviewed before we generalise it to include spin degrees of freedom in section \ref{Fermionic}. In section \ref{FiniteT} we also carry out the first analysis of the theory beyond the classical limit to explore higher order corrections to the result in \cite{Paul1}. Following this a full quantum theory of interacting worldlines is described and quantised in section \ref{contact}. Some supporting calculations on our regularisation scheme are given in the Appendix. 

\section{Bosonic particles -- the classical electric field}
\label{Bosonic}
We begin by working in $D$ spatial dimensions and consider a static charged particle at position $\mathbf{a}$ and an oppositely charged particle at $\mathbf{b}$. The classical electric dipole field for this configuration is the well known solution to Maxwell's equations in the presence of these point particles. In \cite{Paul1} an alternative proposal was made which generates the electric field by carrying out an average over a set of curves joining the two sources. This concept goes some way to reviving Faraday's notion of electric flux lines and it is this calculation that we now review and extend.

An expression satisfying Gauss' law\footnote{We put $\epsilon_{0} = 1$ and denote the electric charge by $q$ so $\nabla \cdot \mathbf{E}^{\prime} = q\delta^{D}\left(\mathbf{x - a}\right) - q\delta^{D}\left(\mathbf{x - b}\right)$} was given in \cite{Paul1}:
\begin{equation}
	E^{\prime i }\left(\mathbf{x}\right) = q\int_{C} d\tau \, \frac{d\omega^{i}}{d\tau} \delta^{D}\left(\boldsymbol{\omega}\left(\tau\right) -\mathbf{x}\right)
	\label{EField}
\end{equation}
where the integral is taken over any curve C with endpoints at $\mathbf{a}$ and $\mathbf{b}$. The form of this field is reminiscent of the Dirac string which was introduced to describe the field of a magnetic monopole \cite{Diracstring1, Diracstring2} but we shall use it here in the context of electrostatics, where $C$ is interpreted as a single line of flux stretched between $\mathbf{a}$ and $\mathbf{b}$. The expression for $\mathbf{E}^{\prime}$ does not satisfy $\nabla \times \mathbf{E}^{\prime} = 0$, and so cannot represent the physical field of a pair of charged particles, but we shall see that its statistical average does. This average is over all curves with endpoints fixed at $\mathbf{a}$ and $\mathbf{b}$ and is defined in reparameterisation invariant form as
\begin{equation}
	\left<\Omega\left(\boldsymbol{\omega}\right)\right>_{T}\! =\! \frac{1}{\mathcal{Z}} \int_{\boldsymbol{\omega}\left(0\right) = \mathbf{a}}^{\boldsymbol{\omega}\left(1\right) = \mathbf{b}} \mathscr{D}e \mathscr{D}\boldsymbol{\omega} \,\, \Omega\left(\boldsymbol{\omega}\right) \delta\left(\int_{0}^{1} e\, d\tau - T\right) e^{-S\left[\boldsymbol{\omega}, e\right]}\,; \quad S\left[\boldsymbol{\omega}, e\right] = \int_{0}^{1}  \, \frac{\dot{\boldsymbol{\omega}}^{2}}{2e} d\tau
	\label{average}
\end{equation}
where $\Omega$ is any reparameterisation invariant functional of the path (we work in Euclidean space). The action is that of Brink, diVecchia and Howe \cite{BdVH} describing the dynamics of a bosonic point particle (we take the particle to have zero mass\footnote{A mass term could have been included in the action defined in (\ref{average}) via the inclusion of a cosmological term $\frac{1}{2}m_{0}^{2} \int_{0}^{1} e \,d\tau$. Usually it is necessary to do so in order to remove a divergence in the functional determinant from the integral over $\bomega$ by a renormalisation of this bare mass $m_{0}$ to a physical mass $m$. In the current case the divergence is cancelled by $\mathcal{Z}$ and since $E^{\prime}$ does not involve the einbein the effect of including $m$ would also be cancelled by the normalisation.}) whose worldline traces out a curve $\bomega\left(\tau\right)$, which the square of the einbein, $e\left(\tau\right)$, equips with a one dimensional metric. The normalisation constant is defined by $\left<1\right> = 1$.  This action is invariant under diffeomorphisms $\tau \rightarrow\tilde{\tau}$ that preserve the parameter interval if under such reparameterisations $e^2(\tau)$ transforms as a metric and $\bomega\left(\tau\right)$ as a scalar:
\begin{equation}
\tilde{e}(\tilde\tau)\,d\tilde\tau=e(\tau)\,d\tau; \qquad  \tilde{\bomega}(\tilde\tau)={\bomega(\tau)}.
\end{equation}
This idea of averaging a quantity over curves with fixed endpoints is similar in spirit to some applications of the worldline approach to the Casimir effect -- in \cite{Cas1, Cas2} the force between two planar bodies induced by quantum fluctuations of a scalar field was arrived at by counting closed worldlines which intersect both surfaces. Open worldlines have also been used to calculate dressed propagators for matter interacting with a background gauge field \cite{NaserTree}.

The $\delta$-function in (\ref{average}) will appear unfamiliar, both in the context of quantum mechanics and the worldline formalism, since it picks out paths of fixed intrinsic length\footnote{In the worldline approach it would be more common to interpret $T$ as a modulus which remains after gauge fixing the einbein. As has been discussed in \cite{Paul1}, the average defined in (\ref{average}) retains reparameterisation invariance, despite the $\delta$-function restricting the functional integral to the subset of einbeins which give the paths a fixed length.} $T$. The Coulomb field for the pair of particles was arrived at in \cite{Paul1} by taking the average of $\mathbf{E}^{\prime}$ in the limit as $T \rightarrow \infty$:
\begin{equation}
	\lim_{T\rightarrow \infty}\left<\mathbf{E}^{\prime}\left(\mathbf{x}\right)\right>_{T} = \lim_{T\rightarrow \infty}\frac{q}{\mathcal{Z}}\int_{\mathbf{a}}^{\mathbf{b}}\mathscr{D}e\mathscr{D}\bomega\, \delta\left(\int e \,d\tau - T \right) \, \int_{C} d\tau\, \dot{\bomega}\, \delta^{3}\left(\bomega\left(\tau\right) - \mathbf{x}\right)\,e^{-S\left[\bomega, e\right]}.
\end{equation}
Taking this limit means that the lines of flux can fluctuate on a macroscopic scale. This is necessary because the field at the point $\mathbf{x}$ is constructed by adding together the contributions from all curves that start and end at the location of the charges and which intersect the spatial point in question. In order that the field permeate all of space, rather than being confined to regions that are of a comparable scale to the distance between the points (in comparison to $\sqrt{T}$), we must allow the worldlines unconstrained fluctuations. For comparison to \cite{Us1, Us2}, which involved averages based on Polyakov's formulation of spinning strings, $T$ plays a similar role to the string tension, $\alpha^{\prime}$. In our earlier work we found it necessary to take the limit in which the strings were tensionless ($\alpha^{\prime} \rightarrow \infty$) which allowed the fluctuations of the string worldsheet to be macroscopically large. This ensured that the averages were not dominated by the minimal spanning surface of the string boundaries and allowed an Abelian gauge field to be supported throughout space-time. In the current context, the large $T$ limit plays an analogous role, ensuring that the field lines have access to all of space. In section \ref{FiniteT} we will return to discuss the consequences of finite values of $T$, exploring also the low $T$ limit for both bosonic and fermionic particles.

In the current context of point particle worldlines, it is also possible to interpret $T$ as an effective temperature so that the functional average defined by (\ref{average}) can be regarded as a thermal statistical average. This was explained in \cite{Paul1} and can be seen by exploiting the reparameterisation invariance to set the einbein to be constant along the worldline. The constraint then imposes $e = T$ and the action can be written as $\beta \mathcal{H}$ where $\beta = T^{-1}$ and $\mathcal{H} = \int_{0}^{1} \dot{\bomega}^{2} / 2 d\tau$. Then integrating over all curves with fixed endpoints can be seen as allowing thermal fluctuations of the worldlines, the relative weighting of each configuration being related to the temperature. As discussed above, desiring that large-scale fluctuations are not penalised in the average requires the limit $T \rightarrow \infty$. We will return to this interpretation at the end of this section when we consider the effect of applying our theory in Minkowski space via a Wick rotation of one of the spatial coordinates. In the meantime we think of the constraint in the geometrical context of holding the length measured along the worldline constant and now return to demonstrate that the average of $\mathbf{E}^{\prime}$ results in the Coulomb field.  

In \cite{Paul1} this result was shown using techniques derived from canonical quantisation. Here we confirm the result using the functional methods we will use for the fermionic generalisation of this claim. To do so we must address the overcounting caused by the reparameterisation symmetry. Following Polaykov \cite{polyB} (see also \cite{PaulStrings, Us2}) we recall that for open curves the measure on the space of one dimensional metrics can be written 
\begin{equation}
\mathscr{D}e = dc \mathscr{D}V\sqrt{\left(\int d\tau\, e\right)\textrm{Det}\left(D^{\dagger} D\right)}\,; \qquad DV = \frac{d}{d\tau}\left(V e\right)
\end{equation}
where $c$ represents a scaling of the einbein and $V$ generates an infinitesimal reparameterisation. This volume element follows because any metric can be written as a combined scaling, generated by $c$, plus reparameterisation, generated by $V$, about some reference metric, $\hat{e}$. We choose to expand about $\hat{e} = \top$, a constant, whereby \newline $D^{\dagger}DV = -e^{-2}\frac{d}{d\tau}\left(e^{-1} \frac{d}{d\tau}\left(e V\right)\right) = -\frac{1}{\top^{2}}\frac{d^{2}V}{d\tau^{2}}$. The infinite product over the eigenvalues of this operator can be $\zeta$-function regulated as in Appendix A, and the volume element becomes
\begin{equation}
	\mathscr{D}e = d\top \mathscr{D}V,
\end{equation}
whilst the constraint on path lengths becomes $\delta\left(\top - T\right)$. Since the components of the functional average are taken to be invariant under reparameterisations the functional integral with respect to $V$ evaluates to the volume of the reparameterisation group, which cancels with the corresponding contribution from the normalisation constant. So for the average of $\mathbf{E}^{\prime}\left(x\right)$ we must determine
\begin{equation}
	\left<\mathbf{I}\left(\mathbf{x}\right)\right> \equiv \frac{q}{\mathcal{Z}^{\prime}} \int \mathscr{D}\boldsymbol{\omega} \int \frac{d^{D}\mathbf{k}}{\left(2\pi\right)^{D}} \int_{0}^{1} d\tau_{1} \, \frac{d \boldsymbol{\omega}\left(\tau_{1}\right)}{d \tau_{1}} e^{i \mathbf{k} \cdot \left(\boldsymbol{\omega}\left(\tau_{1}\right) - \mathbf{x}\right)} e^{-\int_{0}^{1}  \frac{\boldsymbol{\dot{\omega}}^{2}}{2T}d\tau}
	\label{Efixed}
\end{equation}
where $\mathcal{Z}^{\prime}$ is what remains after the volume of the reparameterisation group has been cancelled from $\mathcal{Z}$ and we have used the Fourier decomposition of the $\delta$-function. The insertion that arises has a familiar form -- it is the one dimensional version of the vertex operator used in bosonic string theory and it frequently appears in calculations in the worldline formalism:
\begin{equation}
	V_{k}^{\mu}\left(\tau\right) = \dot{\omega}^{\mu}\left(\tau\right) e^{i k \cdot \omega\left(\tau\right)}.
	\label{vertex}
\end{equation} 
Unlike in string theory we integrate this operator over all momenta. From the point of view of the one dimensional quantum theory (\ref{Efixed}) is the amplitude for the path from $\mathbf{a}$ to $\mathbf{b}$ to pass through the point $\mathbf{x}$. Before calculating this expectation value we also note that the structure of the insertion allows us to constrain its dependence on momentum -- if we contract (\ref{vertex}) with $k_{\mu}$ and integrate over $\tau$ we find a contribution only from the endpoints of the domain
\begin{equation}
	\int d\tau~ k_{\mu}V_{k}^{\mu}\left(\tau\right) = -i\left(e^{i k \cdot \omega\left(1\right)} - e^{i k \cdot \omega\left(0\right)}\right) = -i\left(e^{i k \cdot b} - e^{-i k \cdot a}\right)
	\label{Gauss1}
\end{equation}
which we shall refer to as the generalised Gauss' law. 

The insertion can be generated by introducing a source, $\mathbf{j}\left(\tau\right)$, and defining $\mathbf{J}\left(\tau\right) = -q\frac{d\mathbf{j}}{d\tau} - i \mathbf{k} \delta\left(\tau - \tau_{1}\right)$. Then the above equation becomes
\begin{equation}
	\left.\frac{1}{\mathcal{Z}^{\prime}} \int \mathscr{D}\boldsymbol{\omega} \int \frac{d^{D}\mathbf{k}}{\left(2\pi\right)^{D}} \int_{0}^{1} d\tau_{1} \, e^{-i \mathbf{k} \cdot \mathbf{x}}\frac{\delta}{\delta \mathbf{j}\left(\tau_{1}\right)} e^{-\int_{0}^{1}  \frac{\boldsymbol{\dot{\omega}}^{2}}{2T} + \boldsymbol{\omega} \cdot \mathbf{J} d\tau}\right|_{\mathbf{j} = \mathbf{0}}.
\end{equation}
We split $\boldsymbol{\omega}$ into its classical part in the absence of a source and a piece which absorbs the source and accounts for the quantum fluctuations $\boldsymbol{\omega}\left(\tau\right) = \boldsymbol{\omega}_{c}\left(\tau\right) + \tilde{\boldsymbol{\omega}}\left(\tau\right)$. Here $\boldsymbol{\omega}$ satisfies the source-free classical equation of motion $\frac{-1}{T}\frac{d^{2}\boldsymbol{\omega}}{d\tau^{2}} = \mathbf{0}$ with endpoints at $\mathbf{a}$ and $\mathbf{b}$:
\begin{equation}
	\boldsymbol{\omega}\left(\tau\right) = \mathbf{a} + \left(\mathbf{b} - \mathbf{a}\right) \tau.
\end{equation}
The Dirichlet boundary conditions mean that $\tilde{\boldsymbol{\omega}}\left(\tau\right)$ is required to vanish at the endpoints. It can be split into a classical piece, $\tilde{\boldsymbol{\omega}}_{c}$, satisfying $\frac{-1}{T} \frac{d^{2}\tilde{\boldsymbol{\omega}}_{c}}{d\tau^{2}} = \mathbf{J}\left(\tau\right)$ and a quantum fluctuation, $\bar{\boldsymbol{\omega}}\left(\tau\right)$. Integrating over $\bar{\boldsymbol{\omega}}$ leads to a functional determinant (which we evaluate with $\zeta$-function regularisation in Appendix A) and because the path is open there are also boundary contributions. We find
\begin{align}
	&\frac{ \left(2\pi T\right)^{-\frac{D}{2}} e^{-\frac{\left(\mathbf{b - a}\right)^{2}}{2T}}} {\mathcal{Z}^{\prime}} \int \frac{d^{D}\mathbf{k}}{\left(2\pi\right)^{D}} \int_{0}^{1} d\tau_{1} \, e^{-i \mathbf{k} \cdot \mathbf{x}}\frac{\delta}{\delta \mathbf{j}\left(\tau_{1}\right)}& \nonumber \\
	&\qquad\exp{\left(-\frac{1}{2} k^{2} G\left(\tau_{1}, \tau_{1}\right) - iq \int_{0}^{1} d\tau \, \mathbf{k} \cdot \mathbf{j}\left(\tau\right) \frac{d}{d \tau} G\left(\tau_{1}, \tau\right)\right.}& \nonumber \\
	&\hphantom{\exp{\left(-\frac{1}{2} k^{2} G\left(\tau_{1}, \tau_{2}\right) \right.}}\left. + \frac{q^{2}}{2} \int_{0}^{1}\!\int_{0}^{1} d\tau d\tau^{\prime}\, \mathbf{j}\left(\tau\right) \cdot \mathbf{j}\left(\tau^{\prime}\right) \frac{d}{d\tau} \frac{d}{d \tau^{\prime}} G\left(\tau, \tau^{\prime}\right) \right) \nonumber \\
	&\qquad\left.\hphantom{\exp{\left(-\frac{1}{2} k^{2} G\left(\tau_{1}, \tau_{2}\right) \right.}}\exp{\left(-q\int_{0}^{1} d\tau \mathbf{j}\left(\tau\right) \cdot \frac{d}{d \tau} \boldsymbol{\omega}_{c} \left(\tau\right) + i \mathbf{k} \cdot \boldsymbol{\omega}_{c}\left(\tau_{1}\right)\right)}\right|_{\mathbf{j} = \mathbf{0}}.
	\label{integrated}
\end{align}
In the above equation $G\left(\tau_{1}, \tau_{2}\right)$ is the Green function on the interval $\left[0, 1\right]$ with Dirichlet boundary conditions $G\left(0, \tau_{2}\right) = 0 = G\left(1, \tau_{2}\right)$. Its explicit form is
\begin{equation}
	G\left(\tau_{1}, \tau_{2}\right) = -\frac{T}{2}\left(\left|\tau_{1} - \tau_{2}\right| - \left(\tau_{1} + \tau_{2}\right) + 2\tau_{1}\tau_{2}\right);\qquad  -\frac{1}{T}\frac{d^{2}G}{d\tau_{1}^{2}} = \delta\left(\tau_{1} - \tau_{2}\right).
	\label{G}
\end{equation}
and $G\left(\tau_{1}, \tau_{1}\right)$ is the coincident limit, which in one dimension is finite. Worldline Green functions play an important role in first quantised formulations of field theory and as such have been analysed extensively \cite{Dai1}. The numerator of the prefactor in (\ref{integrated}) is the heat-kernel for the free particle and this cancels exactly with $\mathcal{Z}^{\prime}$. Carrying out the functional differentiation and setting $\mathbf{j = 0}$ yields for the average in momentum representation
\begin{equation}
	\left<\mathbf{I}\left(\mathbf{k}\right)\right> = -q\int_{0}^{1} d\tau_{1} \left[\boldsymbol{\dot{\omega}}_{c}\left(\tau_{1}\right) - \frac{1}{2} i \mathbf{k} \frac{d}{d \tau_{1}} G\left(\tau_{1}, \tau_{1}\right)\right]e^{-\frac{1}{2} k^{2} G\left(\tau_{1}, \tau_{1}\right)} e^{i \mathbf{k} \cdot \boldsymbol{\omega}_{c}\left(\tau_{1}\right)}.
	\label{Iavr}
\end{equation}
Since we will eventually take the limit $T\rightarrow \infty$ it is useful at this point to extract the $T$ dependence in order to make an expansion in powers of $\frac{1}{T}$. Define then $\tilde{G}\left(\tau_{1}, \tau_{2}\right) \equiv \frac{1}{T} G\left(\tau_{1}, \tau_{2}\right)$ and note that 
\begin{equation}
	\tilde{G}\left(\tau_{1}, \tau_{1}\right) = -\tau_{1} \left(\tau_{1} - 1\right).
	\label{G0b}
\end{equation}
This function vanishes on the boundary and is increasing (decreasing) for $\tau < 1/2$ ($\tau > 1/2)$. In this parameterisation (\ref{Iavr}) becomes
\begin{equation}
	\left<\mathbf{I}\left(\mathbf{k}\right)\right> = -q\int_{0}^{1} d\tau_{1} \left[\boldsymbol{\dot{\omega}}_{c}\left(\tau_{1}\right) - \frac{T}{2} i \mathbf{k} \frac{d}{d \tau_{1}} \tilde{G}\left(\tau_{1}, \tau_{1}\right)\right]e^{-\frac{1}{2} k^{2} T \tilde{G}\left(\tau_{1}, \tau_{1}\right)} e^{i \mathbf{k} \cdot \boldsymbol{\omega}_{c}\left(\tau_{1}\right)}.
	\label{IavrG}
\end{equation}
In the large $T$ limit the suppression caused by the exponent $\exp{\left(- \frac{1}{2} k^{2} T \tilde{G}\left(\tau_{1}, \tau_{1}\right) \right)}$ causes the contributions to the integral to arise primarily when $\tau_{1} \rightarrow 0$ and $\tau_{1} \rightarrow 1$ (where the coincident Green function vanishes). We therefore expand the field $\boldsymbol{\omega}$ about these points and integrate a small distance, $h$, along the worldline. At lowest order in $\frac{1}{T}$ we have:
\begin{align}
	&-q\int_{0}^{h} d\tau_{1}\, \left[\boldsymbol{\dot{\omega}}_{c}\left(0\right) - \frac{T}{2} i \mathbf{k} \frac{d}{d \tau_{1}} \tilde{G}\left(\tau_{1}, \tau_{1}\right)\right]e^{-\frac{1}{2} k^{2} T \tilde{G}\left(\tau_{1}, \tau_{1}\right)} e^{i \mathbf{k} \cdot \boldsymbol{\omega}_{c}\left(0\right)} \nonumber \\
	&-q\int_{1-h}^{1} \!\!d\tau_{1} \left[\boldsymbol{\dot{\omega}}_{c}\left(1\right) - \frac{T}{2} i \mathbf{k} \frac{d}{d \tau_{1}} \tilde{G}\left(\tau_{1}, \tau_{1}\right)\right]e^{-\frac{1}{2} k^{2} T \tilde{G}\left(\tau_{1}, \tau_{1}\right)} e^{i \mathbf{k} \cdot \boldsymbol{\omega}_{c}\left(1\right)} 
	\label{highT}
\end{align}
and the damping caused by the form of the coincident Green function in the exponent allows the integration regions to be extended by setting $h = \frac{1}{2}$ as we take $T \rightarrow \infty$. Each integral has two terms, the second of which provides the leading order contribution:
\begin{align}
	-\frac{T i \mathbf{k}}{2} \int_{0}^{\frac{1}{2}} d\tau_{1}\, \frac{d}{d\tau_{1}} \tilde{G}\left(\tau_{1}, \tau_{1}\right)e^{-\frac{1}{2} k^{2} T \tilde{G}\left(\tau_{1}, \tau_{1}\right)} &= \frac{ i \mathbf{k}}{k^{2}}\left[e^{-\frac{1}{2}k^{2} T \tilde{G}\left(\frac{1}{2}, \frac{1}{2}\right)} - 1\right] \nonumber \\
	&\rightarrow -\frac{ i \mathbf{k}}{k^{2}}
\end{align}
where the last line holds as $k^{2} T \rightarrow \infty$. Noting that for $0 \le \tau \le 1/2$ we have $\tilde{G}\left(\tau, \tau\right) \ge \frac{\tau}{2}$ the first terms of each line in (\ref{highT}) can be bounded
\begin{align}
	\int_{0}^{\frac{1}{2}} d\tau_{1}\, e^{ -\frac{1}{2}k^{2}T\tilde{G}\left(\tau_{1}, \tau_{1}\right)} &\le \int_{0}^{\frac{1}{2}} d\tau_{1} \,e^{-\frac{1}{4}k^{2}T\tau_{1}} \nonumber \\
	&= \frac{4} {k^{2} T} \left[1-e^{-\frac{1}{8}k^{2}T} \right]
\end{align}
which is $\mathcal{O}\left(\frac{1}{k^{2}T}\right)$. Putting this together with the contribution from the other end of the path we arrive at the momentum space expression
\begin{equation}
	\left<\mathbf{I}\left(\mathbf{k}\right)\right> = \frac{q i \mathbf{k}}{k^{2}}\left(e^{i \mathbf{k} \cdot \mathbf{a}} - e^{i \mathbf{k} \cdot \mathbf{b}}\right) + \mathcal{O}\left(\frac{1}{k^{2}T} \right)
\end{equation}
We check our answer by compatibility with the generalised Gauss' law: dotting with $\mathbf{k}$ produces the expected contribution (\ref{Gauss1}). Finally, taking the limit $T \rightarrow \infty$ and setting $D = 3$ the result can be written in position space as 
\begin{equation}
	\left<\mathbf{I}\left(\mathbf{x}\right)\right>  = \frac{q}{4\pi}\boldsymbol{\nabla}\left( -\frac{1}{ \left|\mathbf{x} - \mathbf{a}\right| } +  \frac{1}{ \left|\mathbf{x} - \mathbf{b}\right| } \right)
\end{equation}
which is indeed the classical dipole electric field. This average therefore determines $F_{0i}$ for static oppositely charged particles. In \cite{Paul1} the case of magnetostatics was also considered for a fixed closed current carrying wire, and the time varying situation was also examined. These cases require treating the curve $C$ as dynamical so that the natural weight becomes not the action of a point particle but that of extended objects, naturally leading to the use of string theory. This picture was explored further in \cite{Us1, Us2} where a theory of tensionless spinning strings interacting on contact was shown to be equivalent to the Abelian theory of quantum electrodynamics.

The calculation presented above provides an interesting method of determining the classical static dipole electric field, albeit somewhat unconventional. The utility of the functional approach is the ease with which it can be extended. Before generalising to fermionic particles we show that a simple change can instead generate the classical electric field due to a static point particle. This is desirable since it is presumably more useful to deal with a single particle rather than be constrained to dealing with oppositely charged pairs (except when considering the worldlines of virtual particle / anti-particle pairs). For a single particle at the point $\mathbf{a}$ we proceed as above with the exception that we constrain only one end of the worldlines. This is a simple change of the boundary condition at the upper end of the interval; the variation of the action shows that the only other consistent choice we can make is the Neumann condition $\dot{\boldsymbol{\omega}}\left(1\right) = 0$.

There are three effects of this change. Firstly the determinant of the kinetic operator $\frac{-1}{T}\frac{d^{2}}{d\tau^{2}}$ becomes independent of $T$ (see Appendix A). This has no relevance because it is cancelled by the same change in $\mathcal{Z^{\prime}}$. Of more importance is the change in the Green function. With the new boundary conditions we find
\begin{equation}
	G^{\prime}\left(\tau_{1}, \tau_{2}\right) = -\frac{T}{2}\left(\left|\tau_{1} - \tau_{2}\right| - \left(\tau_{1} + \tau_{2}\right)\right)\,; \qquad G^{\prime}\left(\tau_{1}, \tau_{1}\right) = T \tau_{1}.
\end{equation}
Note that the coincident Green function now only vanishes at $\tau_{1} = 0$, the end of the curve tied to the particle at $\mathbf{a}$. Finally the classical solution must clearly differ; now $\boldsymbol{\omega}_{c} = \mathbf{a}$ satisfies the equation of motion and boundary conditions. In particular $\dot{\boldsymbol{\omega}} = 0$ and there are no boundary contributions from the classical action so that (\ref{IavrG}) becomes
\begin{equation}
	\left<\mathbf{I}\left(\mathbf{k}\right)\right> = q\int_{0}^{1} d\tau_{1}  \frac{1}{2} i \mathbf{k} \frac{d}{d \tau_{1}} G^{\prime}\left(\tau_{1}, \tau_{1}\right)e^{-\frac{1}{2} k^{2} G^{\prime}\left(\tau_{1}, \tau_{1}\right)} e^{i \mathbf{k} \cdot \mathbf{a}}.
\end{equation}
Defining $G^{\prime}\left(\tau_{1}, \tau_{2}\right) \equiv T\tilde{G}^{\prime}\left(\tau_{1}, \tau_{1}\right)$ we could again consider the above equation for large $T$ whereby the integrand is suppressed by the exponent $\exp{\left(-\frac{1}{2}k^{2}T \tilde{G}^{\prime}\left(\tau_{1}, \tau_{1}\right)\right)}$ except for $\tau_{1} \approx 0$. The integrand is a total derivative however so we can determine the exact answer. We obtain
\begin{equation}
	\left<\mathbf{I}\left(\mathbf{k}\right)\right> = \frac{ i q\mathbf{k}}{k^{2}}\left(1 - e^{-\frac{1}{2}k^{2} T}\right) e^{i \mathbf{k} \cdot \mathbf{a}}
	\label{I1T}
\end{equation}
which in the limit $T \rightarrow \infty$ has inverse Fourier transform equal to the static electric field of a point particle of charge $q$ at the spatial point $\mathbf{a}$. 
\begin{equation}
	\left<\mathbf{I}\left(\mathbf{x}\right)\right> = \frac{q}{4\pi}\frac{\mathbf{x - a}}{ \left|\mathbf{x} - \mathbf{a}\right|^{3}}.
\end{equation}
It is interesting to note that (\ref{I1T}) gives the exact average at finite $T$. We will come to explore corrections to the classical fields for both of the above configurations of particles below. We have already commented on the calculation of the static magnetic field of a \textit{closed} loop of current carrying wire given in \cite{Paul1}. This required a functional average over surfaces with boundary fixed to the wire and already invoked the use of string theory. There the two dimensional worldsheet Green function was required to vanish at both ends of the string since the endpoints were constrained to lie on the boundary. The analogous method of applying Dirichlet boundary conditions to only one end corresponds to opening up the wire into a small segment and ensures that the only contribution to the average comes from a strip close to the end of the string fixed to the wire. The average then yields the Biot-Savart law for that segment of wire. Mixed boundary conditions are discussed further in the full interacting string theory in Appendix A of \cite{Us2} where it is shown to provide propagators of the field theory.

For the most part we took $D$ to be arbitrary, only specialising to $D = 3$ spatial dimensions for the sake of compatibility with \cite{Paul1} and an illustration of some of the physical content of the average. In a four dimensional space-time we will have to deal with 4 bosonic coordinates so we append $\omega^{0}$ to the three fields $\bomega$ considered above. The Euclidean average is then constructed over all paths which intersect the spatial point $x^{\mu} = \left(x^{0}, \mathbf{x}\right)$ whose endpoints are fixed to the particles at $a^{\mu} = \left(a^{0}, \mathbf{a}\right)$ and $b^{\mu} = \left(b^{0}, \mathbf{b}\right).$ The average takes the form
\begin{equation}
	\left<I^{\mu}\left(x\right)\right>= \left<q\int d\tau \frac{d \omega^{\mu}}{d \tau} \delta^{4}\left(\omega\left(\tau\right) - x\right)\right>.
\end{equation}
If we restrict attention to the case of a single point particle at position $a^{\mu}$ discussed above the generalisation of (\ref{I1T}) provides the Euclidean space average
\begin{equation}
	\left<I^{\mu}\left(k\right)\right> = \frac{i qk}{k^{2}}\left(1 - e^{-\frac{1}{2}k^{2}T}\right)e^{i k \cdot a}.
\end{equation}
We can interpret this in Minkowski space by treating the boundary data $a^{\mu}$ as some fixed point on the worldline of the particle and Wick rotating the above result. In the $T \rightarrow \infty$ limit then
\begin{equation}
	\left<I_{\mu}\left(x\right)\right> = \frac{q}{4\pi^{2}} \partial_{\mu} \frac{1}{\left|x - a\right|^{2} + i\epsilon}
	\label{Imink}
\end{equation}
where we use the $i\epsilon$ procedure to specify the positions of the poles. 

The physical interpretation of this result is less obvious because it relies on the choice of $a^{\mu}$ (also $b^{\mu}$ had we included a second particle). If we return to a static picture then consider an observer at $x^{\mu}$ in the rest frame of the charged particle. The calculation in $D = 3$ spatial dimensions was an average over all possible particle paths starting at $\mathbf{a}$ and passing through $\mathbf{x}$ but in four dimensional space-time they are also required to pass though $\mathbf{x}$ at the time $x^{0} = t$, say. We are restricting our attention to a static configuration and suppose that we ought to integrate over all possible starting points on the worldline $a^{\mu}$ at which the path $\omega^{\mu}$ could begin whilst still passing through $x^{\mu}$. Recall that the parameter $T$ can be interpreted as a temperature, which in Euclidean space implies that our calculation is a thermal average. In section 4 of \cite{Paul1} an argument was given based on the construction of thermal Green functions that the retarded solutions to Maxwell's equations are inherited from the Feynman propagator if the Minkowski space calculation is seen as a finite temperature quantum expectation value. Application of this procedure to the current problem provides
\begin{equation}
	\left<I_{\mu}\left(x\right)\right> = \frac{q}{4\pi} \partial_{\mu} \frac{\delta\left(x^{0} - a^{0} - \left|\mathbf{x} - \mathbf{a}\right|\right)} {\left|\mathbf{x} - \mathbf{a}\right|}
	\label{ICausal}
\end{equation}
so that the only contribution comes from paths whose endpoints are joined along the causal light-cone. In the rest frame of the particle its worldline has constant $\mathbf{a}$ and integrating (\ref{ICausal}) with respect to $a^{0}$ yields the static electric field expected at $x^{0} = t$:
\begin{equation}
	F_{00} = \left<I_{0}\left(x\right)\right> = 0; \qquad F_{0i} = \left<I_{i}\left(x\right)\right> = \frac{q}{4\pi}\frac{\mathbf{x - a}}{\left|\mathbf{x - a}\right|^{3}}
	\label{speed}
\end{equation} 
The same result could be arrived at by integrating (\ref{Imink}) over $a^{0}$ immediately if the contour for the $a^{0}$ integral is chosen to fall above the poles on the real axis; the discussion of the thermal average in \cite{Paul1} can thus be seen as justification for this choice of contour. The result (\ref{speed}) is then familiar as the field created at position $(x^{0}, \mathbf{x})$ due to a point electric charge which had  position $\mathbf{a}$ a time $\left|\mathbf{x - a}\right|$ earlier, which appears in the more familiar   action at a distance theories discussed in the introduction.

It should be stressed that this procedure yields the correct static field but is not applicable for the general time dependent problem. This has been dealt with in \cite{Paul1} and \cite{Us1}. It requires string theory to correctly describe the dynamics of extended curves whose boundaries are fixed to the worldlines of the charged particles. The contact interaction is then between fundamental strings, between worldsheets rather than worldlines. The static problem considered above is an unusual way to arrive at the electric field but is nonetheless of interest because of its straightforward generalisation. In the following sections we first include spin degrees of freedom before returning to an analysis of the result when the parameter $T$ is taken to be finite. We then generalise the work to form a full quantum theory of point particles with contact interactions.

Following on from the bosonic theory we wish to provide an extension to the theory in \cite{Paul1}. The following sections comprise the new contribution in this article. We have three aims in sight. The first is to extend the work to fermionic fields and the second is to determine the corrections to both results which are present at finite $T$. We finally ask whether the interaction exponentiates as in \cite{Us1, Us2} -- we shall address this issue in section \ref{contact}.
\section{Fermionic particles}
\label{Fermionic}
In this section we consider a theory of spin 1/2 particles and generalise the theory presented above for application to this case. Spinning particles have been considered in action at a distance theories in the past \cite{spin1, spin2, spin3} and it is natural to expect some spin-dependent modifications to the results found in the previous section. We continue to work with massless particles for simplicity and it will also prove most convenient to work in four dimensional Euclidean space (this is more natural for fermionic theories -- we discuss how we may relate this to the three dimensional case below). To deal with fermions it is useful to construct the theory in superspace. We therefore introduce the Grassmann variable $\theta$ to extend the parameter domain $\tau \rightarrow \left(\tau, \theta\right)$. We further introduce the scalar superfield\footnote{The strange looking factors of $e^{\frac{1}{2}}$ are necessary for the action given in the text to reduce to the familiar action of Brink, Di-Veccia and Howe \cite{BdVH}. Another convention for the superfields exists where we scale $\psi \rightarrow e^{-\frac{1}{2}}\psi$ and $\chi \rightarrow e^{\frac{1}{2}}\chi$.}
\begin{equation}
	\mathbf{X}\left(\tau, \theta\right) = \omega\left(\tau\right) + \theta e^{\frac{1}{2}}\left(\tau\right)\psi \left(\tau\right)
\end{equation}
and the supereinbein
\begin{equation}
	\mathbf{E}\left(\tau, \theta\right) = e\left(\tau\right) + \theta e^{\frac{1}{2}}\left(\tau\right) \chi \left(\tau\right),
\end{equation}
where $\psi$ is the superpartner to $\omega$ and $\chi$ is the gravitino. We also define the superderivative
\begin{equation}
	D = \partial_{\theta} + \theta \partial_{\tau}.
\end{equation}
Under the local supersymmetry transformations parameterised by $V\left(\tau\right)$, the generator of reparameterisations, and $\eta\left(\tau\right)$, a Grassmann function generating pure supersymmetry transformations,
\begin{equation}
\tau \rightarrow \tau + V\left(\tau\right) + \theta \eta\left(\tau\right); \qquad \theta \rightarrow \theta + \eta\left(\tau\right) + \frac{1}{2}\theta \dot{V}\left(\tau\right)
\end{equation}
the superderivative transforms homogeneously
\begin{equation}
	D \mathbf{X} \rightarrow \Lambda\left(\tau, \theta\right) D \mathbf{X}
\end{equation}
and the supereinbein transforms as
\begin{equation}
	\mathbf{E} \rightarrow \Lambda^{2}\left(\tau, \theta\right) \mathbf{E}
\end{equation}
where $\Lambda\left(\tau, \theta\right) = 1 + \frac{1}{2}\dot{V}\left(\tau\right) + \theta \dot{\eta}\left(\tau\right)$. Requiring the integration measure to transform as $d\tau d\theta \rightarrow \Lambda^{-1}\left(\tau, \theta\right) d\tau d\theta$ the following action is invariant:
\begin{equation}
	S\left[\mathbf{E}, \mathbf{X}\right] = \frac{1}{2}\int d\tau d\theta \, \mathbf{E}^{-1} D^{2}\mathbf{X} \cdot D\mathbf{X}.
	\label{Ssusy}
\end{equation}
Integrating over $\theta$ allows this to be cast in the more familiar component form\footnote{The supersymmetry transformations of the component fields follow from those of the superfields and (in the absence of reparameterisations) are: $\delta_{\eta} \omega = \eta \psi; ~ \delta_{\eta} \psi = \frac{\eta}{e}\left( \dot{\omega} - \frac{1}{2} \chi \psi\right);~ \delta_{\eta}e = \eta \chi;~ \delta_{\eta} \chi = 2\dot{\eta}$. Under reparameterisations, $\psi$ is a worldline scalar like $\omega$, whilst $\chi$ transforms as $e$.}
\begin{equation}
	\frac{1}{2}\int d\tau \, e^{-1} \dot{\omega}^{2} + \dot{\psi}\cdot \psi - \frac{\chi}{e} \dot{\omega} \cdot \psi.
	\label{Scpt}
\end{equation}
Under canonical quantisation the equations of motion for the auxiliary fields $\chi$ and $e$ lead to the first class constraints $p \cdot \psi = 0$ and then $p^{2} = 0$ respectively. Here $p^{\mu} = \dot{\omega}^{\mu} / e$ is the momentum corresponding to the field $\omega$. The fundamental anti-commutation relations $\left\{\psi^{\mu}, \psi^{\nu}\right\} = \delta^{\mu\nu}$ can be solved by taking $\psi^{\mu} = \frac{1}{\sqrt{2}}\gamma^{\mu}$ so that on the state space the former of the constraints enforces the Dirac equation on physical states and the latter informs us the particle has zero mass \cite{BdVH}. Below we shall pursue again the functional quantisation of this action.

We also require the supersymmetric generalisation of the interaction and the constraint on the intrinsic length of the path. The natural invariant interaction term is
\begin{equation}
	\mathbf{I}\left(x\right) = q\int d\tau d\theta \, D\mathbf{X}\, \delta^{4}\left(\mathbf{X} - x\right),
	\label{Isusy}
\end{equation}
whose functional average will be taken over all worldlines with endpoints attached to fixed charges. Fourier decomposing the $\delta-$function and integrating over $\theta$ puts this into the form
\begin{equation}
	q\int \frac{d^{4} k}{\left(2\pi\right)^{4}} \int d\tau \left(\dot{\omega}- e\psi ik \cdot \psi\right)e^{i k \cdot \left(\omega - x\right)}
	\label{Icpt}
\end{equation}
which is analogous to the supersymmetric vertex operator familiar in the context of the spinning string and the worldline formulation of spinor field theory:
\begin{equation}
	V_{k}^{\mu}\left(\tau\right) = \left(\dot{\omega}^{\mu}\left(\tau\right) - e\left(\tau\right) \psi^{\mu}\left(\tau\right) i k \cdot \psi\left(\tau\right)\right) e^{i k \cdot \omega\left(\tau\right)}
\end{equation} 
Such an expression was examined in a string setting in \cite{Us1, Us2} but for now we continue to explore the point particle theory. We contend that the generalisation of the electric field part of the field strength tensor should be generated by a functional average of (\ref{Icpt}) with an appropriate weight. This weight will be of course that corresponding to the action in (\ref{Scpt}). Note also that the anti-commuting nature of $\psi$ ensures that the generalised Gauss' law (\ref{Gauss1}) still holds for this interaction.

The final element we need is the supersymmetric version of the constraint on path lengths. Previously we inserted $\delta\left(\int e \,d\tau - T\right)$ into the functional average but this changes under supersymmetry transformations. Introducing an arbitrary Grassmann number $\Xi$, the natural invariant quantity is
\begin{equation}
	\delta\left(\int d\tau d\theta \, \mathbf{E}^{\frac{1}{2}} - \frac{1}{2}\Xi\right)
\end{equation}
which imposes a Grassmannian constraint on $\chi$ rather than on the metric. In the massless case considered here it is not possible to construct a local function of $e$ and $\chi$ which is supersymmetric. Instead we follow Polyakov \cite{polyB} and give a superspace version of the non-local and invariant quantity he termed the superlength:
\begin{equation}
	\delta\left(-\frac{1}{2} \int\!\int d\tau_{1}d\theta_{1}\, \mathbf{E}^{\frac{1}{2}}\left(\tau_{1}, \theta_{1}\right) D_{1}D_{2} G\left(\tau_{1}, \theta_{1}; \tau_{2}, \theta_{2}\right) \mathbf{E}^{\frac{1}{2}} \left(\tau_{2}, \theta_{2}\right) d\tau_{2}d\theta_{2}\, - T \right),
\end{equation}
where $G\left(\tau_{1}, \theta_{1}; \tau_{2}, \theta_{2}\right) = \left|\tau_{1} - \tau_{2} - \theta_{1}\theta_{2}\right|$ is a superinvariant Green function and $D_{i}$ is the super-derivative acting on the parameters $\left(\tau_{i}, \theta_{i}\right)$. In components the first of these constraints takes the form
\begin{equation}
	\delta\left(\frac{1}{2} \int d\tau \, \chi\left(\tau\right) - \frac{1}{2}\Xi\right)
	\label{chi}
\end{equation}
whilst the latter can be written
\begin{equation}
	\delta\left(\int d\tau \, e\left(\tau\right) - \frac{1}{8}\int\!\int d\tau_{1} \chi\left(\tau_{1}\right) \textrm{sg}\left(\tau_{1} - \tau_{2}\right)\chi\left(\tau_{2}\right)d\tau_{2}\, - T\right)
	\label{superlength}
\end{equation} 
with $\textrm{sg}\left(\tau\right) = \frac{\tau}{\left|\tau\right|}$ equal to the sign of its argument. We shall require the superlength constraint in the functional average because it will be seen to provide the appropriate fixing of the einbein and we will also impose the complementary constraint (\ref{chi}) which will similarly fix $\chi$.

To continue the calculation it is necessary to determine the measure on the space of the gravitino and $\psi$ and also to specify the boundary conditions we will use. It is well known that the purpose of the fields $\psi$ is to represent the $\gamma$ matrices (in canonical quantisation we have seen it suffices to take $\psi^{\mu} \propto \gamma^{\mu}$), which is why it is desirable to work in a four dimensional space-time. For instance, consider the partition function of a free fermion $\psi$ coupled to a source $\zeta$. Integrating the exponential of the action over $\psi$ yields an object with spinor indices and choosing the boundary conditions on the integral allows us to extract each component of the result. Specifically in Appendix C of \cite{Us2} we have shown that, for example,
\begin{equation}
	\left. \int \mathscr{D}\psi \,e^{-\int d\tau \left(\frac{1}{2} \dot{\psi} \cdot \psi + \zeta \cdot \psi \right)}\right|_{\psi_{2} = i\psi_{1}; \psi_{4} = -i\psi_{3}}^{\psi_{2} = -i\psi_{1}; \psi_{4} = i\psi_{3}} = \mathscr{T}\left(e^{-\frac{1}{\sqrt{2}}\int d\tau \,\zeta \cdot \gamma}\right)_{11}\,,
	\label{gamma}
\end{equation}
where $\mathscr{T}$ represents time ordering along the worldline. We also showed that the volume element for $\chi$ can be written
\begin{equation}
	\mathscr{D}\chi = d\chi_{0}\mathscr{D}\eta \left( \int e^{-1} d\tau  \right)^{-\frac{1}{2}} \textrm{Det}^{-\frac{1}{2}} \left( -\left(\frac{1}{e} \frac{d}{d \tau}\right)^{2} \right)
\end{equation}
where $\chi_{0}$ is the constant piece of $\chi$ required for consistency with the boundary conditions of open worldlines. This volume elements follows because $\chi$ can be expressed as a supersymmetry transformation, generated by $\eta$, plus a change proportional to $e$ about some reference, $\hat{\chi}$. We gauge fix by expanding about $\hat{e} = \top$ -- a constant -- and $\hat{\chi} = \chi_{0}$.

These are consistent choices which cover the physically distinct configurations and on this gauge slice the action becomes
\begin{equation}
	\frac{1}{2}\int d\tau\, \frac{\dot{\omega}^{2}}{\top} + \dot{\psi}\cdot \psi - \frac{\chi_{0}}{\top} \dot{\omega} \cdot \psi
\end{equation}
whilst the interaction vertex is given by
\begin{equation}
	\left(\dot{\omega}^{\mu}\left(\tau\right) - \top \psi^{\mu}\left(\tau\right) i k \cdot \psi\left(\tau\right)\right)e^{i k \cdot \omega \left(\tau\right)}.
\end{equation}
Similarly, using $\zeta$-function regularisation for the functional determinants (see Appendix A), the volume elements become
\begin{equation}
	\mathscr{D}e\mathscr{D}\chi =  d\top d\chi_{0} \mathscr{D}V \mathscr{D}\eta.
\end{equation}
The super-length constraint reduces to
\begin{equation}
	\delta\left(\top - T\right),
\end{equation}
and the analogous version for $\chi$ becomes
\begin{equation}
	\delta\left(\chi_{0} - \Xi\right),
\end{equation}
which can be used to carry out the integrals over $\top$ and $\chi_{0}$. We shall again eventually take $T$ to infinity so we can expand in powers of $\frac{1}{T}$ and we will take the dimensionful Grassmann parameter $\Xi$ to vanish. The action and insertions are locally supersymmetric so that the integrals with respect to $V$ and $\eta$ will evaluate to the volumes of the reparameterisation group and supersymmetry group respectively. These constants, albeit formally divergent, are cancelled by their counterparts if we normalise against the bare partition function. 

We pause here to derive an important result using (\ref{gamma}). We consider the expectation value over the fermionic fields $\left<\psi^{\mu}\left(\tau\right) \psi^{\nu}\left(\tau\right)\right>\big|_{\alpha\beta}$ where we have attached the spinor indices $\alpha$ and $\beta$ to the beginning and end of the worldline respectively\footnote{We have written the fields at equal times but there is of course a time-ordering issue. We take the convention that the field to the left is understood to be evaluated at an infinitesimally greater time, $\epsilon$, than that on the right, after which we take the limit $\epsilon \rightarrow 0^{+}$.}. The insertions can be generated by introducing a fermionic source $\eta$ and carrying out functional differentiation, after which we set $\eta = \frac{\chi_{0}}{2\sqrt{2}T}\dot{\omega}$ to produce the linear term in the action. Then (\ref{gamma}) gives 
\begin{equation}
	 \left<\psi^{\mu}\left(\tau\right) \psi^{\nu}\left(\tau\right)\right>_{\alpha\beta} =\frac{1}{2\mathcal{Z}}\left. \int d\chi_{0}\delta\left(\chi_{0} - \Xi\right) \frac{\delta}{\delta \eta^{\mu}\left(t\right)} \frac{\delta}{\delta \eta^{\nu}\left(t\right)}\mathscr{T}\left(e^{\int d\tau \,\eta \cdot \gamma}\right)_{\alpha\beta} \right|_{\eta = \frac{ \chi_{0} \dot{\omega}}{2\sqrt{2}T}}
\end{equation}
where the constant $\mathcal{Z}$ is determined in the appendix. If we impose $\Xi = 0$ the result of the functional differentiation and the integral over $\chi_{0}$ is
\begin{equation}
	\left<\psi^{\mu}\left(\tau\right) \psi^{\nu}\left(\tau\right)\right> = \frac{1}{2}\left(\delta^{\mu\nu} - \gamma^{\mu}\gamma^{\nu}\right)
\end{equation}
which crucially is independent of the field $\omega$, since by setting $\Xi = 0$ we have decoupled the fields $\omega$ and $\psi$ in the action. This result is consistent with the symmetry properties of the Grassman fields and will play a key role in the forthcoming calculations.

We now return to (\ref{Icpt}) and take its functional average over worldlines fixed to charges at $a^{\mu}$ and $b^{\mu}$. Carrying out the integral over $\top$ we are left with the gauge fixed Fourier space expectation value 
\begin{align}
	\left<I^{\mu}\left(k\right)\right>\! &=\! \frac{q}{\mathcal{Z}}\int \!\mathscr{D}\mathbf{\omega} \mathscr{D}\psi d\chi_{0} \delta\left(\chi_{0} - \Xi\right)\int_{0}^{1} d\tau_{1} \left(\dot{\omega}^{\mu}\! -\! T\psi^{\mu} i k \cdot \mathbf{\psi}\right)e^{i k \cdot \mathbf{\omega}} e^{-\int_{0}^{1} d\tau \frac{\mathbf{\dot{\omega}}^{2}}{2T} + \frac{1}{2}\mathbf{\dot{\psi}} \cdot \mathbf{\psi} - \frac{\chi_{0}}{2T} \mathbf{\dot{\omega}} \cdot \mathbf{\psi}} \nonumber \\
	&= \frac{q}{\mathcal{Z^{\prime}}} \int \mathscr{D}\omega \int_{0}^{1} d\tau_{1} \left(\dot{\omega}^{\mu} - \frac{Ti}{2}\left( k^{\mu} - \gamma^{\mu}k\cdot\gamma\right)\right)e^{i k \cdot \mathbf{\omega}}e^{-\int_{0}^{1} d\tau \frac{\mathbf{\dot{\omega}}^{2}}{2T}}
\end{align}
where for the last line we have integrated over $\psi$ to produce the gamma matrices and integrated over $\chi_{0}$ before putting $\Xi = 0$. It remains to determine the expectation values of these integrands and carry out the integral over $\tau_{1}$. The integral over $\omega$ can be found to give
\begin{equation}
 q\int_{0}^{1} d\tau_{1} \left(\dot{\omega}_{c}^{\mu} \left(\tau_{1}\right) - \frac{Ti}{2}\left( k^{\mu} - \gamma^{\mu}k\cdot\gamma\right) - \frac{1}{2} i k^{\mu} \frac{\partial}{\partial \tau_{1}} G\left(\tau_{1}, \tau_{1}\right) \right) e^{-\frac{1}{2}k^{2} G\left(\tau_{1}, \tau_{1}\right)} e^{i k\cdot \omega_{c} \left(\tau_{1}\right)}
\end{equation}
Making further use of the redefinition $G\left(\tau_{1}, \tau_{2}\right) = T \tilde{G}\left(\tau_{1}, \tau_{2}\right)$ this can be cast in the form
\begin{equation}
q	\int_{0}^{1} d\tau_{1} \left(\dot{\omega}_{c}^{\mu}\left(\tau_{1}\right) -\frac{Ti}{2}\left( k^{\mu} - \gamma^{\mu}k\cdot\gamma\right) - \frac{T}{2} i k^{\mu} \frac{\partial}{\partial \tau_{1}} \tilde{G}\left(\tau_{1}, \tau_{1}\right) \right) e^{-\frac{1}{2}k^{2}T \tilde{G}\left(\tau_{1}, \tau_{1}\right)} e^{i k \cdot \omega_{c} \left(\tau_{1}\right)}
\label{omegaf}
\end{equation}
From the exponent $\exp{\left(-\frac{1}{2} k^{2} T \tilde{G}\left(\tau_{1}, \tau_{1} \right) \right)}$ it is easy to understand that the latter two terms in brackets will contribute at leading order in $\frac{1}{T}$ and that the first term will once again be sub-leading. In the limit of large $T$ we can follow the calculations of the bosonic theory and approximate the integral by evaluating the classical path $\omega_{c}$ on each boundary and then integrating a short distance, $h$, along the worldline. We calculate
\begin{align}
	&q\int_{0}^{h} d\tau_{1} \left( \dot{\omega}_{c}^{\mu} \left(0\right) - \frac{Ti}{2}\left( k^{\mu} - \gamma^{\mu}k\cdot\gamma\right) - \frac{T}{2} i k^{\mu} \frac{\partial}{\partial \tau_{1}} \tilde{G}\left(\tau_{1}, \tau_{1}\right) \right) e^{-\frac{1}{2}k^{2}T \tilde{G}\left(\tau_{1}, \tau_{1}\right)} e^{i k \cdot \omega_{c} \left(0\right)} \nonumber \\
	+ &q\int_{1 - h}^{1}d\tau_{1} \left( \dot{\omega}_{c}^{\mu} \left(1\right) - \frac{Ti}{2}\left( k^{\mu} - \gamma^{\mu}k\cdot\gamma\right) - \frac{T}{2} i k^{\mu} \frac{\partial}{\partial \tau_{1}} \tilde{G}\left(\tau_{1}, \tau_{1}\right) \right) e^{-\frac{1}{2}k^{2}T \tilde{G}\left(\tau_{1}, \tau_{1}\right)} e^{i k \cdot \omega_{c} \left(1\right)} 
\end{align}
and as before the suppression caused by the exponent at large $T$ allows us to extend the integrands by setting $h = 1/2$. Carrying out the integrals in the large $T$ limit leads to 
\begin{align}
	\left<I^{\mu}\left(k\right)\right> = q\left[ \frac{ i }{k^{2}  } \gamma^{\mu} k \cdot \gamma \left( e^{i k \cdot a} - e^{i k \cdot b}\right) + \mathcal{O}\left(\frac{1}{k^{4}T}\right)\right].
\end{align}
In the limit $T \rightarrow \infty$ only the first term contributes and so we have determined
\begin{equation}
	\left<I\left(k\right)\right> = \frac{ iq }{k^{2} }\, \gamma\, k \cdot \gamma  \left(e^{i k \cdot a} - e^{i k \cdot b}\right).
	\label{Ifermion}
\end{equation}
We may double check our work by contracting with $k_{\mu}$ to verify against (\ref{Gauss1}) that the index structure is correct. In position space the expression above becomes
\begin{equation}
	\left<I\left(x\right)\right> = \frac{q}{4\pi^{2}}\gamma \,\gamma \cdot \nabla  \left(\frac{1}{\left|x - a\right|^{2}} - \frac{1}{\left|x - b\right|^{2}} \right),
	\end{equation}
which we take as the generalisation of the Coulomb field found in the previous section. 

We conclude this section by commenting on how this approach could be applied to a three dimensional space. The main obstacle to this case is the structure of (\ref{gamma}) which imposes boundary conditions relating pairs of the $\psi^{\mu}$. Furthermore the proof of this equation presented in \cite{Us2} does not hold for the three dimensional version of the Clifford algebra because the chirality operator does not anti-commute with the other matrices; in particular the trace of a product of an odd number of gamma matrices does not in general vanish. We present two alternatives to address this issue. The first is to use the symmetry of the problem to choose a coordinate system such that the two charges are placed on the $z = 0$ plane and to restrict our attention to determine the field only on this plane. Then we need consider a super-field, $\boldsymbol{\omega}$, which consists of only two components and the fields $\psi$ essentially become the $\sigma$-matrices $\sigma_{1}$ and $\sigma_{2}$:
\begin{align}
\left<I^{i}\left(\mathbf{k}\right)\right> &= \frac{q}{\mathcal{Z}}\int \mathscr{D}\boldsymbol{\omega} \mathscr{D}\psi d\chi_{0} \int_{0}^{1} d\tau_{1} \left(\dot{\omega}^{i} - T\psi^{i} i \mathbf{k} \cdot \boldsymbol{\psi}\right)e^{i \mathbf{k} \cdot \boldsymbol{\omega}} e^{-\int_{0}^{1} d\tau \frac{\boldsymbol{\dot{\omega}}^{2}}{2T} + \boldsymbol{\dot{\psi}} \cdot \boldsymbol{\psi} + \frac{\chi_{0}}{T} \boldsymbol{\dot{\omega}} \cdot \boldsymbol{\psi}} \nonumber \\
	&= \frac{q}{\mathcal{Z}^{\prime}} \int \mathscr{D}\boldsymbol{\omega} \int_{0}^{1} d\tau_{1} \left(\dot{\omega}^{i} - \frac{Ti}{2}\left(k^{i} - \sigma^{i}  \mathbf{k} \cdot \boldsymbol{\sigma}\right)\right)e^{i \mathbf{k} \cdot \boldsymbol{\omega}} e^{-\int_{0}^{1} d\tau \frac{\boldsymbol{\dot{\omega}}^{2}}{2T}} \nonumber \\
\end{align}
where $i \in \left\{1, 2\right\}$. Carrying out the integral over the two-dimensional field $\boldsymbol{\omega}$ and following the same steps as the four dimensional case leads to
\begin{equation}
	\left<\mathbf{I}\left(\mathbf{k}\right)\right> = \frac{ iq}{k^{2} } \boldsymbol{\sigma}\, \mathbf{k} \cdot \boldsymbol{\sigma} \left(e^{i \mathbf{k} \cdot \mathbf{a}} - e^{i \mathbf{k} \cdot \mathbf{b}}\right).
	\label{I2}
\end{equation}

It is more satisfactory to instead modify the four dimensional theory in such a way that statements can be made for the three dimensional case. This can be done in two equivalent ways. The three dimensional version of the vertex operator is
\begin{equation}
	\mathbf{V}_{k}= \left(\boldsymbol{\dot{\omega}} - e\boldsymbol{\psi} i \mathbf{k}\cdot \boldsymbol{\psi}\right)e^{i \mathbf{k} \cdot \boldsymbol{\omega}}.
\end{equation}
We could use this as an insertion in the four dimensional theory by defining in some inertial frame $I^{\mu} = \left(0, \mathbf{I}\right)$ where the 3-vector $\mathbf{I}$ is constructed out of the three dimensional vertex operator above. Then (\ref{gamma}) continues to hold except that the part of the expectation value which does not cancel with $\mathcal{Z}$ contains only a three dimensional source so the right hand side of that equation becomes
\begin{equation}
	\mathscr{T}\left(e^{-\frac{1}{\sqrt{2}}\int d\tau \boldsymbol{\zeta} \cdot \boldsymbol{\gamma}}\right).
\end{equation}
This is equivalent to beginning with an entirely three dimensional theory but introducing a further pair of fields $\omega_{0}$, $\psi_{0}$. Into the integral of some functional of the three dimensional fields we introduce a supersymmetric invariant factor in the numerator and denominator as follows:
\begin{equation}
	\frac{1}{\mathcal{Z}_{3}}\!\int \!\mathscr{D}\boldsymbol{\omega} \mathscr{D}\boldsymbol{\psi} d\chi_{0} \Omega\left(\boldsymbol{\omega}, \boldsymbol{\psi}\right) e^{-\int_{0}^{1}d\tau\, \frac{\boldsymbol{\dot{\omega}}}{2T} + \frac{1}{2}\dot{\boldsymbol{\psi}} \cdot \boldsymbol{\psi} - \frac{\chi_{0}}{2T} \dot{\boldsymbol{\omega}} \cdot \boldsymbol{\psi}}  \frac{\int \mathscr{D}\omega_{0} \mathscr{D}\psi_{0}   e^{-\int_{0}^{1}d\tau\, \frac{\dot{\omega}_{0}}{2T} + \frac{1}{2}\dot{\psi}_{0} \cdot \psi_{0} - \frac{\chi_{0}}{2T} \dot{\omega}_{0} \cdot \psi_{0}}}{\int \mathscr{D}\omega_{0} \mathscr{D}\psi_{0} e^{-\int_{0}^{1}d\tau\, \frac{\dot{\omega}_{0}}{2T} + \frac{1}{2}\dot{\psi}_{0} \cdot \psi_{0} - \frac{\chi_{0}}{2T} \dot{\omega}_{0} \cdot \psi_{0}}}.   
	\label{one}    
\end{equation}
We then combine the denominator of the fraction with the three dimensional normalisation $\mathcal{Z}_{3}$ to form
\begin{equation}
	\int \mathscr{D}\,\omega \mathscr{D}\psi d\tilde{\chi}_{0}~e^{-\int_{0}^{1}d\tau\, \frac{\dot{\omega}}{2T} + \frac{1}{2}\dot{\psi} \cdot \psi - \frac{\tilde{\chi}_{0}}{2T}\boldsymbol{\dot{\omega}} \cdot \boldsymbol{\psi} - \frac{\chi_{0}}{2T} \dot{\omega}_{0} \cdot \psi_{0} }
	\label{denom}
\end{equation}
where the integration is now over four bosonic and four fermionic fields. We apply the boundary conditions for open paths to both the numerator and denominator of (\ref{one}). The integration in (\ref{denom}) produces an expression dependent on $\chi_{0}$ which feeds back into (\ref{one}). Carrying out the integral over $\chi_{0}$ and the four integrations with respect to $\psi$ produces two terms which conspire to yield
\begin{equation}
	\frac{1}{\mathcal{Z}^{\prime}}\int \mathscr{D}\omega ~ \Omega \left(\boldsymbol{\omega}, \boldsymbol{\gamma}\right) e^{-\int_{0}^{1} d\tau \frac{\dot{\omega}^{2}}{2T}}
\end{equation}
where the normalisation $\mathcal{Z}^{\prime}$ is precisely the correct factor to ensure that $\left<1\right> = 1$. This is what we would calculate if we were to take the expectation of $I^{\mu} = \left(0, \mathbf{I}\right)$ as defined above in a theory with four pairs of fields. It is easy to modify (\ref{Ifermion}) to respect this change:
\begin{equation}
	\left<\mathbf{I}\left(\mathbf{k}\right)\right> = \frac{ iq }{k^{2} } \boldsymbol{\gamma} \, \mathbf{k} \cdot \boldsymbol{\gamma}\left(e^{i \mathbf{k} \cdot \mathbf{a}} - e^{i\mathbf{k} \cdot \mathbf{b}}\right)
\end{equation}
which is to be integrated with respect to the three-vector $\mathbf{k}$. We may choose the representation
\begin{equation}
\gamma^i=\left( \begin{array}{cc}
0 &   -i\sigma^i\\
i\sigma^i& 0  \end{array} \right),
\end{equation}
to show that this approach involves the Pauli matrices in a similar way to (\ref{I2}), but now we need four-index spinors. With this representation $\boldsymbol{\gamma} \, \mathbf{k} \cdot \boldsymbol{\gamma}$ is block diagonal with the two blocks taking the same form as (\ref{I2}), differing from one another by a sign.
\section{Analysis at finite $T$}
\label{FiniteT}
The classical fields of the previous sections were found in the limit that the dimensionful parameter $T$ was taken to be large compared to momenta. We discussed above how this limit can be interpreted as making the intrinsic length of the worldlines stretched between the charged particles macroscopically large. An interesting question is to ask about the form of the statistical average for finite $T$ in order to explore how the fields change. In this section we give the subleading correction to the fields at large $T$ and also consider the opposing limit $T \rightarrow 0$. We shall do so for the lowest order interaction which generates the classical fields. The geometrical interpretation of changing $T$ is to explore the contribution to the average from worldlines of different fixed lengths, but as we have discussed earlier it may also be considered as varying the temperature at which the thermal average is calculated. Therefore, at lower values of $T$, we expect that large fluctuations of the worldlines will be penalised and so the functional averages will become dominated by curves close to the shortest path between the charges.  

\subsection{Corrections in the bosonic case}
We first analyse the bosonic particle with mixed boundary conditions whose high $T$ limit produced the classical electric field of a point particle (\ref{I1T}). That equation was exact in $T$ and its position space form is found by carrying out the inverse Fourier transform. We shall specialise here to the three physical spatial dimensions. Aligning the z-axis parallel to the spatial separation $\mathbf{x - a}$ the angular integrals of (\ref{I1T}) can be carried out to produce
\begin{align}
	\left<\mathbf{I}\left(\mathbf{x}\right)\right>_{T} &=-q\boldsymbol{\nabla}\int_{0}^{\infty}\frac{dk}{\left(2\pi\right)^{2}} 2\frac{\sin{\left(k \left|\mathbf{x - a}\right|\right)}}{k\left|\mathbf{x - a}\right|}\left(1 -e^{-\frac{1}{2}k^{2}T}\right) \nonumber \\
	&=-q\boldsymbol{\nabla} \frac{1}{4\pi \left|\mathbf{x - a}\right|} \left(1 - \textrm{Erf}\left(\sqrt{\frac{\left|\mathbf{x - a}\right|^{2}}{2T}}\right)\right).
	\label{PointT}
\end{align}
At large $T$ this can be expanded in powers of $\frac{1}{T}$ and we find 
\begin{equation}
	\left<\mathbf{I}\left(\mathbf{x}\right)\right>_{T} = -\boldsymbol{\nabla}\frac{q}{4\pi} \left(\frac{1}{\left|\mathbf{x - a}\right|} - 2\sqrt{\frac{1}{2\pi T}} + \frac{1}{3}\sqrt{\frac{1}{2\pi T} } \frac{\left|\mathbf{x - a}\right|^{2}}{T} + \mathcal{O}\left(T^{-\frac{5}{2}}\right) \right)
\end{equation}
valid for $\left|\mathbf{x - a}\right|^{2} / T \ll 1$. The leading order correction arises from the third term in brackets:
\begin{equation}
	\left<\mathbf{I}\left(\mathbf{x}\right)\right>_{T} = \frac{q}{4\pi}\frac{\mathbf{x - a}}{\left|\mathbf{x - a}\right|^{3}} - \frac{q}{6 \pi} \sqrt{\frac{1}{2\pi } }\frac{\mathbf{x - a}}{T^{\frac{3}{2}}} +  \mathcal{O}\left(T^{-\frac{5}{2}}\right)
\end{equation}
giving the finite $T$ deviation from the inverse square law. 

The other limit of interest is at small $T$ so we consider an expansion of (\ref{PointT}) about $T = 0$. The change in functional form is more dramatic because we find
\begin{equation}
	\left<\mathbf{I}\left(\mathbf{x}\right)\right>_{T} = -\frac{q}{2}\boldsymbol{\nabla} \frac{1}{\sqrt{2\pi}}\left(\frac{\sqrt{T}}{\left|\mathbf{x - a}\right|^{2}} - \frac{T^{\frac{3}{2}}}{\left|\mathbf{x - a}\right|^{4}} + \mathcal{O}\left(T^{\frac{5}{2}}\right) \right)
\end{equation}
which carries a different power of the spatial separation. Carrying out the differentiation we acquire
\begin{equation}
	\left<\mathbf{I}\left(\mathbf{x}\right)\right>_{T} = \frac{q}{4\pi}\sqrt{2\pi}\left(\frac{2\sqrt{T}\left(\mathbf{x - a}\right)} {\left|\mathbf{x - a}\right|^{4}} - \frac{4T^{\frac{3}{2}}\left(\mathbf{x - a}\right)}{\left|\mathbf{x - a}\right|^{6}} + \mathcal{O}\left(T^{\frac{5}{2}}\right)\right)
\end{equation}
in the limit that $\left|\mathbf{x - a}\right|^{2} / T \gg 1$. 

\begin{figure}[h]
	\centering
	\includegraphics[width = 0.4 \textwidth]{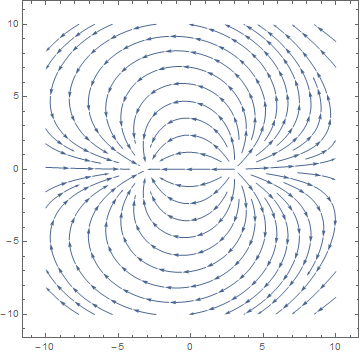}
	\caption{The field lines for large but finite $T$ demonstrating the small deviation from the classical dipole field. In this plot we set $T = 10\left|\mathbf{b - a}\right|^{2}$.}
	\label{figDipole}
\end{figure}

The corrections to the dipole field are less trivial to determine because of the time dependence in $\boldsymbol{\omega}_{c}$. For large $T$ the integrand is still dominated by contributions at either end of the curve. We first expand about $\tau = 0$:
\begin{equation}
	e^{i \mathbf{k}\cdot \boldsymbol{\omega}_{c} \left(\tau_{1}\right)} = e^{i \mathbf{k}\cdot \mathbf{a}}\left(1 + i \mathbf{k} \cdot \left(\mathbf{b - a}\right) \tau_{1} - \frac{k^{2}}{2} \left(\mathbf{k} \cdot\left(\mathbf{b - a}\right)\right)^{2}\tau_{1}^{2} + \ldots\right).
	\label{expan}
\end{equation} 
When integrated against $\exp{\left(-\frac{1}{2}k^{2}T G\left(\tau_{1}, \tau_{1}\right)\right)}$ each extra power of $\tau_{1}$ results in a further power of $\frac{1}{k^{2}T}$. The first term in the square brackets of (\ref{IavrG}) is already subleading in $\frac{1}{T}$ so its contribution to the leading order correction comes from the first term in (\ref{expan}). Integrating this from the boundary to $\tau_{1} = \frac{1}{2}$ gives
\begin{equation}
	\frac{-2q\left(\mathbf{b - a}\right)}{k^{2} T} e^{i \mathbf{k} \cdot \mathbf{a}} + \mathcal{O}\left(T^{-2}\right)
	\label{knoindex}
\end{equation}
plus corrections exponentially suppressed at large $k^{2} T$. At first order the second term of (\ref{IavrG}) provided the classical dipole field and its leading order correction arises from the $\mathcal{O}\left(\tau_{1}\right)$ term in (\ref{expan}). Straightforward integration of this evaluates to
\begin{equation}
	\frac{2 q \mathbf{k}\, \mathbf{k} \cdot \left(\mathbf{b - a}\right)}{k^{4}T} e^{i \mathbf{k} \cdot \mathbf{a}} + \mathcal{O}\left(T^{-2}\right)
	\label{kindex}
\end{equation}
up to further terms which are again exponentially suppressed. It is instructive to combine these two terms as follows: the $i$th component is given by
\begin{equation}
	-\frac{2q}{T} \left[\frac{\delta^{ij}}{k^{2}} - \frac{k^{i}k^{j}}{k^{4}}\right]\left(b - a\right)^{j}e^{i \mathbf{k} \cdot \mathbf{a}}
\end{equation} 
which highlights the transverse nature of the correction. The index structure of (\ref{kindex}) means the integral with respect to $\mathbf{k}$ has the form
\begin{equation}
	\int \frac{d^{3}\mathbf{k}}{\left(2\pi\right)^{3}} \frac{k^{i} k^{j}}{k^{4}} e^{i \mathbf{k} \cdot \left(\mathbf{a-x}\right)} = A \delta^{ij} + B \frac{\left(a - x\right)^{i} \left(a-x\right)^{j}}{\left|\mathbf{a - x}\right|^{2}}
	\label{kanswer}
\end{equation}
for some constants $A$ and $B$ of dimension $\left[\textrm{length}\right]^{-1}$. They can be determined by contracting each side of the above equation first with $\delta_{ij}$ and also with $\left(a - x\right)_{i}\left(a - x\right)_{j}$. The first of these gives
\begin{equation}
	3A + B = \int \frac{d^{3}\mathbf{k}}{\left(2\pi\right)^{3}} \frac{e^{i \mathbf{k} \cdot \left(\mathbf{a-x}\right)}}{k^{2}}  = \frac{-1}{4\pi \left|\mathbf{x} - \mathbf{a}\right|} 
\end{equation}
and the second yields
\begin{equation}
	\left|a - x\right|^{2}\left(A + B\right) = \int \frac{d^{3}\mathbf{k}} {\left(2\pi\right)^{3}} \frac{\left( \mathbf{k} \cdot \left(\mathbf{a - x} \right)\right)^{2} } {k^{4}} e^{i \mathbf{k} \cdot \left(\mathbf{a-x}\right)}.
\end{equation}
Choosing the $z$-axis to align with $\mathbf{x - a}$ again allows the angular integrals to be done and we find that the remaining integral with respect to the magnitude of $\mathbf{k}$ is proportional to
\begin{equation}
	\frac{1} {\left(2\pi\right)^{2}} \int_{0}^{\infty} dk\, \left[\frac{\sin{\left( k \left|\mathbf{x - a}\right|\right)}} {k \left|\mathbf{x - a}\right|} + \frac{2 \cos{\left(k \left|\mathbf{x - a}\right|\right)}} {\left(k \left|\mathbf{x - a}\right|\right)^{2}} - \frac{2 \sin{\left(k \left|\mathbf{x - a}\right|\right)}} {\left(k \left|\mathbf{x - a}\right|\right)^{3}}\right]
\end{equation}
Integrating the second term in square brackets by parts once and the final term by parts twice serves to cancel the first term. It is then easy to check that the integral evaluates to zero so $A + B = 0$ and (\ref{kanswer}) evaluates to
\begin{equation}
	\frac{-1}{8\pi \left|\mathbf{x - a}\right|}\left[\delta^{ij} - \frac{\left(a - x\right)^{i} \left(a-x\right)^{j}}{\left|\mathbf{a - x}\right|^{2}}\right].
\end{equation}
So (\ref{kindex}) evaluates to
\begin{equation}
	\frac{q}{4\pi T \left|\mathbf{x - a}\right|}\left[\left(\mathbf{b - a}\right) + \left(\mathbf{x - a}\right)\frac{ \left(\mathbf{x - a}\right)\cdot\left(\mathbf{b - a}\right)}{\left|\mathbf{x - a}\right|^{2}}\right] + \mathcal{O}\left(T^{-2}\right),
\end{equation}
to which it remains to add the contribution with the analogous calculation for the other end of the curve fixed to the point $\mathbf{b}$. The final correction at order $\frac{1}{T}$ is determined to be
\begin{align}
	&\frac{q}{4\pi T \left|\mathbf{x - a}\right|}\left[\left(\mathbf{b - a}\right) + \left(\mathbf{x - a}\right)\frac{ \left(\mathbf{x - a}\right)\cdot\left(\mathbf{b - a}\right)}{\left|\mathbf{x - a}\right|^{2}}\right] \nonumber \\
	-& \frac{q}{4\pi T \left|\mathbf{x - b}\right|}\left[\left(\mathbf{b - a}\right) + \left(\mathbf{x - b}\right)\frac{ \left(\mathbf{x - b}\right)\cdot\left(\mathbf{b - a}\right)}{\left|\mathbf{x - b}\right|^{2}}\right]
\end{align}
and is easy to check that this is divergence free. This is the variation from the classical dipole field for large values of $T$ which is present in this model. It is interesting to note that with the relaxation of the limit to large but finite $T$ comes dependence on the direction $\left(\mathbf{b} - \mathbf{a}\right)$ which is independent of the spatial point in question. In Fig \ref{figDipole} we provide an example of the deviation from the well-known dipole field for finite $T$ by plotting the streamlines of the electric field.

We finally turn to the low $T$ limit of the dipole field for which it is more convenient to carry out the integral with respect to $\mathbf{k}$ of (\ref{IavrG}), before looking at an expansion in powers of $T$. Here we shall see more striking dependence on the direction of separation between the two charges, since as $T\rightarrow 0$ one expects the fluctuations of the worldlines to be almost completely frozen out. Then the field will be concentrated in a small region (measured with respect to $T$) about the straight line from $\mathbf{a}$ to $\mathbf{b}$. The result of carrying out the inverse Fourier transform of (\ref{IavrG}) is
\begin{equation}
	\left< \mathbf{I}\left(\mathbf{x}\right) \right>_{T}\! = \!\frac{-q} {\left(2\pi\right)^{\frac{3}{2}}}\int_{0}^{1}  \frac{d\tau_{1}}{\left(T \tilde{G}\left(\tau_{1}, \tau_{1}\right)\right)^{\frac{3}{2}}} \left[ \frac{\dot{\tilde{G}}\left(\tau_{1}, \tau_{1}\right)} {2\tilde{G}\left(\tau_{1}, \tau_{1}\right)} \left(\mathbf{x} - \boldsymbol{\omega}_{c}\right) - \dot{\boldsymbol{\omega}}_{c}\right]\exp{\left(-\frac{\left(\mathbf{x} - \boldsymbol{\omega}_{c}\right)^{2}} {2 T \tilde{G}\left(\tau_{1}, \tau_{1}\right)}\right)},
	\label{lowT}
\end{equation}
Examining the $T$-dependence of this expression we see that the limit $T \rightarrow 0$ provides a representation of the $\delta$ function so that the field is supported only on the classical straight line path joining $\mathbf{a}$ to $\mathbf{b}$:
\begin{align}
	&\delta^{3}\left(\int_{0}^{1} d\tau_{1} \,\left( \mathbf{x} - \boldsymbol{\omega}_{c}\left(\tau_{1}\right)\right) \right) \nonumber \\
	=&\,\delta^{3}\left(\int_{0}^{1} d\tau_{1} \,\left( \mathbf{x} - \left(\mathbf{a} + \left(\mathbf{b - a}\right)\tau_{1}\right) \right) \right).
\end{align}
This makes sense since the $T \rightarrow 0$ limit implies the worldline's intrinsic length is small, so it must find the configuration that minimises the intrinsic distance traced out by the particle. Alternatively, as we have stated above, there can be no thermal fluctuation about the classical solution to the equations of motion.

To determine the form of the field at finite $T$ we shall employ a Laplace approximation. For small $T$ the contribution to the integrand of (\ref{lowT}) is concentrated about the positions of the maxima of the exponent $\exp{\left(\frac{\left(\mathbf{x} - \boldsymbol{\omega}\left(\tau_{1}\right)\right)^{2}}{2 T \tilde{G}\left(\tau_{1}, \tau_{1}\right)}\right)}$. The precise version of Laplace's method we shall invoke is that for small $T$ and arbitrary well-behaved functions $f\left(\tau\right)$ and $g\left(\tau\right)$
\begin{equation}
	\int_{0}^{1} d\tau f\left(\tau\right) e^{-\frac{1}{T} g\left(\tau\right)} = \sum_{\tau_{0}} \sqrt{\frac{2\pi T}{\ddot{g} \left(\tau_{0}\right) } }f\left(\tau_{0}\right) e^{-\frac{1}{T}g\left(\tau_{0}\right)}\left(1 + \mathcal{O}\left(T\right)\right)
	\label{laplace}
\end{equation}
where the $\tau_{0}\in \left[0, 1\right]$ are determined by the condition that $g\left(\tau_{0}\right)$ be a maximum. In this case a straightforward calculation shows that the exponent of (\ref{lowT}) attains a single maximum within the integration range at\footnote{This expression is consistent with the behaviour of the system under exchange of $\mathbf{a}$ and $\mathbf{b}$ because this is equivalent to sending $\tau \rightarrow 1 - \tau$.}
\begin{equation}
	\tau_{0} = \frac{\left|\mathbf{x - a}\right|}{\left|\mathbf{x - a}\right| + \left|\mathbf{x - b}\right|}.
	\label{tau0}
\end{equation}

If the spatial point $\mathbf{x}$ lies on the line joining $\mathbf{a}$ to $\mathbf{b}$ then the exponent vanishes at this value of $\tau_{1}$, in agreement with the $T \rightarrow 0$ limit which provides $\delta$-function support on this line. Furthermore the first term in square brackets of (\ref{lowT}) vanishes because $\mathbf{x} - \boldsymbol{\omega}_{c}\left(\tau_{0}\right) = 0$. In this case the field is approximated at lowest order in $T$ by
\begin{align}
	&\frac{q}{2\pi T \tilde{G}\left(\tau_{0}, \tau_{0}\right)} \frac{\left(\mathbf{b - a}\right)}{\left|\mathbf{b - a}\right|} \nonumber \\
	=\,&\frac{q}{2\pi T} \frac{\left|\mathbf{b - a}\right|}{\left|\mathbf{x - a}\right|\left|\mathbf{x - b}\right|} \left(\mathbf{b - a}\right)
\end{align}
which has its minimum half way between the charges, where its magnitude is
\begin{equation}
	\frac{2q}{\pi T}
\end{equation}

\begin{figure}[t]
\centering
	\subfloat[][The magnitude of the electric field]
	{
		\includegraphics[width = 0.4 \textwidth]{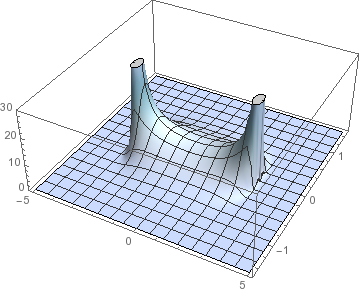}
	}
	\subfloat[][The stream lines of the electric field]
	{
		\includegraphics[width = 0.4 \textwidth]{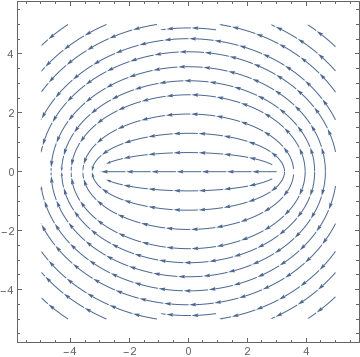}
	}
	\caption{Field magnitude and field streamlines in the low $T$ limit -- in this plot $T = \frac{1}{60}\left|\mathbf{b - a}\right|^{2}$.}
	\label{figE1}
\end{figure}

\begin{figure}[t!]
\centering
	\subfloat[][The magnitude of the electric field]
	{
		\includegraphics[width = 0.4 \textwidth]{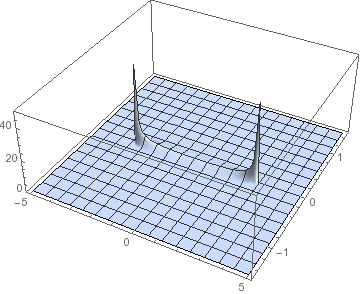}
	}
	\subfloat[][The stream lines of the electric field]
	{
		\includegraphics[width = 0.4 \textwidth]{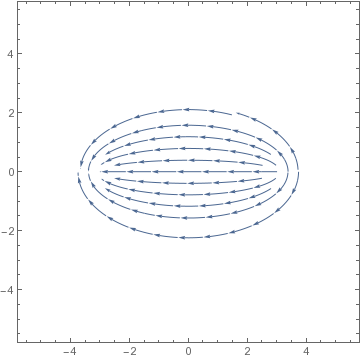}
	}	
	\caption{Field magnitude and field streamlines in the low $T$ limit -- in this plot $T = \frac{1}{1000} \left|\mathbf{b - a}\right|^{2}$. The field decays exponentially as described in the text which gives rise to the white-space in the plot, in which the field is negligibly small.}
	\label{figE2}
\end{figure}

Away from this line the exponent in (\ref{laplace}) enforces an exponential decay in the magnitude of the field. Indeed we find
\begin{align}
	\frac{\left(\mathbf{x} - \boldsymbol{\omega}_{c}\left(\tau_{0}\right)\right)^{2}} {2  \tilde{G}\left(\tau_{0}, \tau_{0}\right)} = \frac{1}{2}&\left[\vphantom{\frac{\left|\mathbf{x - a}\right| + \left|\mathbf{x - b}\right|}{\left|\mathbf{x - b}\right|}}\left|\mathbf{x - a}\right|\left|\mathbf{x - b}\right|\left(\left|\mathbf{x- a}\right| + \left|\mathbf{x - b}\right|\right)^{2} \right. \nonumber \\
	&\left.- 2 \left(\mathbf{x - a}\right)\cdot \left(\mathbf{x - b}\right) \frac{\left|\mathbf{x - a}\right| + \left|\mathbf{x - b}\right|}{\left|\mathbf{x - b}\right|} + \left|\mathbf{x - a}\right|\left|\mathbf{x - b}\right|\right]
	\label{decay}
\end{align} 
and the direction of the field depends on the spatial point through the first term in (\ref{lowT}). It is possible to use (\ref{laplace}) to determine the contribution at an arbitrary spatial point but the result is less illuminating than a visual representation of the field lines and the field magnitude. It is most useful to plot the streamlines, tangent to the field at each spatial point. Fig. \ref{figE1} and Fig \ref{figE2} show the field strength and direction on the $z = 0$ plane of a pair of oppositely charged particles placed at positions $\mathbf{a} = \left(3, 0, 0\right)$ and $\mathbf{b} = \left(-3,0 ,0\right)$ for two values of the parameter $T$. We have imposed a sharp cut-off about the positions of the charges to avoid the divergence encountered there.

\begin{figure}[h]
\centering
	\includegraphics[width = 0.4 \textwidth]{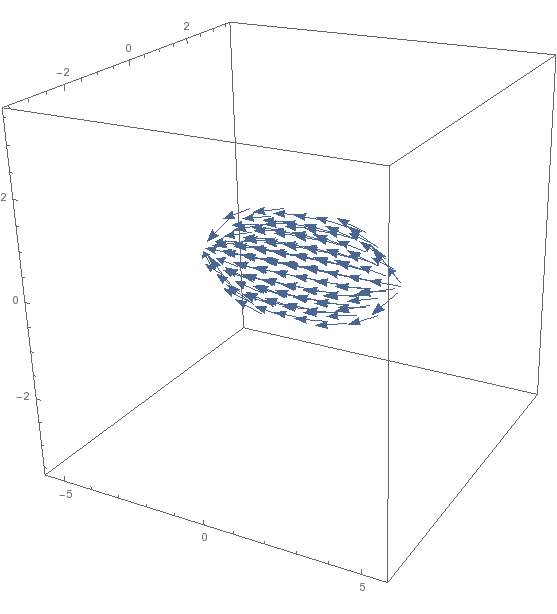}
	\caption{Confining field lines in the low $T$ limit -- here $T = \frac{1}{500}\left|\mathbf{b - a}\right|^{2}$.}
	\label{figE3}
\end{figure}

The form of these field configurations suggests that the low $T$ limit of this theory gives some sort of confining field, albeit not one with a potential linear in the separation of the charges. In this way $T$ interpolates between the classical field associated to a pair of charges and a regime in which field lines are concentrated about the line joining the charges. The emergence of confining behaviour has briefly been considered for action at a distance theories before \cite{con1, con2} and we add to the discussion with our determination of the field lines with this new model. The three dimensional plot in Fig. \ref{figE3} highlights how the low $T$ lines of flux are compressed into a thin tube. This completes our discussion of the correction to the bosonic theory at finite $T$.

\subsection{Corrections in the fermionic case}
In this case we shall consider four dimensional space-time and will briefly state the result of analysing the order $\frac{1}{T}$ corrections to (\ref{Ifermion}). The general approach is identical to that of the bosonic theory. We have already determined the momentum space correction to the first and last terms in (\ref{omegaf}) in the previous section. For the contribution corresponding to the end of the curve at $a^{\mu}$ the $\mu$th component is given by
\begin{equation}
	\frac{2q}{T} \left[\frac{\delta^{\mu \nu}}{k^{2}} - \frac{k^{\mu}k^{\nu}}{k^{4}}\right]\left(b - a\right)^{\nu}e^{i k \cdot a}.
	\label{f1}
\end{equation} 
We also find a subleading contribution from the middle term in (\ref{omegaf}) which takes the form
\begin{equation}
	\frac{2q}{T}\left[\delta^{\mu\alpha} - \gamma^{\mu}\gamma^{\alpha}\right]\frac{k^{\alpha}k^{\nu}}{k^{4}} \left(b - a\right)^{\nu}e^{i k \cdot a}
	\label{f2}
\end{equation}
so now we must determine the four dimensional integral
\begin{equation}
	\int \frac{d^{4}k}{\left(2\pi\right)^{4}} \frac{k^{\mu} k^{\nu}}{k^{4}} e^{i k \cdot {a - x}}.
	\label{k4}
\end{equation}
Symmetry dictates it must be equal to $A\delta^{\mu\nu} + B\left(a - x\right)^{\mu}\left(a - x\right)^{\nu} \left|a - x\right|^{-2}$ where the constants $A$ and $B$ have dimension [length]$^{-2}$. Proceeding as before we contract with $\delta^{\mu\nu}$ and $\left(a - x\right)^{\mu}\left(a - x\right)^{\nu}$ to produce two equations. The first gives
\begin{equation}
	4A + B = \int \frac{d^{4}k}{\left(2\pi\right)^{4}} \frac{ e^{i k \cdot \left(a - x\right) }}{k^{2}} = \frac{-1} {4\pi^{2} \left|x-a\right|^{2}}
\end{equation}
and the second leads to
\begin{equation}
	|a - x|^{2}\left(A + B\right) = \int \frac{d^{4}k}{\left(2\pi\right)^{4}} \frac{\left(k\cdot \left(a - x\right)\right)^{2} }{k^{4}} e^{i k \cdot \left(a - x\right)}.
\end{equation}
This integral can be carried out by first doing the integral with respect to $k_{0}$, where double poles are encountered at $k_{0} = \pm i \left|\mathbf{k}\right|$. The remaining integral over the three dimensional vector $\mathbf{k}$ can be done by choosing the $z$-axis to align with $\mathbf{x - a}$ as above. Then we are left to determine
\begin{align}
	\frac{\pi}{2} \int_{0}^{\infty} \frac{dk}{\left(2 \pi\right)^{3}} \int_{-1}^{1}d\left(\cos{\theta}\right) k e^{-k\left|x_{0}\right| + i k \left|\mathbf{x}\right| \cos{\theta} }\bigg[\left|x_{0}\right|^{2}\left(k \left|x_{0}\right|-1\right) - 2 i k \left|\mathbf{x}\right|\left|x_{0}\right|^{2} \cos{\theta} \nonumber \\
	 - \left|\mathbf{x}\right|^{2}\left(k \left|x_{0}\right| + 1\right)\cos^{2}{\theta}\bigg].
\end{align}
Integrating over $\theta$ leaves only
\begin{align}
	\frac{\pi}{2} \int_{0}^{\infty} \frac{dk} {\left(2 \pi\right)^{3}} e^{-k \left|x_{0}\right|}\bigg[\left|x_{0}\right|^{2}\left(k \left|x_{0}\right|-1\right) \frac{\sin{k \left|\mathbf{x}\right|}}{k \left|\mathbf{x}\right|} + 2 \left|x_{0}\right|^{2} \left(\frac{\sin{k \left|\mathbf{x}\right|}}{k \left|\mathbf{x}\right|} - \cos{k \left|\mathbf{x}\right|}\right) \nonumber \\
	- \left|\mathbf{x}\right|^{2} \left( k \left|x_{0}\right| + 1\right) \left( \frac{\sin{k \left|\mathbf{x}\right|}}{k \left|\mathbf{x}\right|} + 2\frac{\cos{k \left|\mathbf{x}\right|}}{k^{2}\left|\mathbf{x}\right|^{2}} - 2\frac{ \sin{k \left|\mathbf{x}\right|}} {k^{3}\left|\mathbf{x}\right|^{3}} \right)  \bigg].
\end{align}
Carrying out the $k$-integral several times by parts yields $\left|x - a\right|^{2} \left(A  + B\right) = \frac{1}{8 \pi^{2}}$ so (\ref{k4}) evaluates to
\begin{equation}
	\frac{-1}{8 \pi^{2} \left|x - a\right|^{2}}\left[\delta^{\mu\nu} - 2\frac{\left(a - x\right)^{\mu}\left(a - x\right){^\nu}}{\left|x - a\right|^{2}}\right].
\end{equation}
So in position space (\ref{f1}) becomes
\begin{equation}
	\frac{-q}{4\pi^{2}T \left|x - a\right|^{2}}\left[\left(b - a\right)^{\mu} + 2\left(x - a\right)^{\mu}\frac{\left(x - a\right)\cdot \left(b - a\right)}{\left|x -a\right|^{2}}\right],
\end{equation}
for which it can be verified that the divergence vanishes, and (\ref{f2}) becomes 
\begin{equation}
	\frac{-q}{4\pi^{2}T\left|x - a\right|^{2}}\left(\delta^{\mu\nu} - \gamma^{\mu}\gamma^{\nu}\right)\left[\left(b - a\right)^{\nu} - 2\left(x - a\right)^{\nu} \frac{\left(x - a\right)\cdot\left(x - a\right)}{\left|x - a\right|^{2}}\right],
\end{equation}
which is also divergence free. The first order correction is found by subtracting these and including the contribution from the other end of the curve. We find at order $\frac{1}{T}$ the expectation value of the insertion evaluates to
\begin{align}
	\left<I^{\mu}\left(x\right)\right>_{T} = &\frac{q}{4\pi^{2}T\left|x - a\right|^{2}}\gamma^{\mu}\,\gamma \cdot \left[\left(b - a\right) - 2\left(x - a\right) \frac{\left(x - a\right)\cdot\left(x - a\right)}{\left|x - a\right|^{2}}\right] \nonumber \\
-&\frac{q}{4\pi^{2}T\left|x - b\right|^{2}}\gamma^{\mu}\,\gamma \cdot \left[\left(b - a\right) - 2\left(x - b\right)^{\nu} \frac{\left(x - b\right)\cdot\left(x - b\right)}{\left|x - b\right|^{2}}\right]. 
\end{align}
The form of this correction has a similar functional form to that of the bosonic particle discussed above. 

We may also ask about the low $T$ expansion to determine the behaviour of the system in this regime. It is again useful to carry out the integral over $k$ first to arrive at
\begin{align}
	\left< I^{\mu} \left(x\right) \right>_{T} = \frac{-q} {\left(2\pi\right)^{2}}\int_{0}^{1}  \frac{d\tau_{1}}{\left(T \tilde{G}\left(\tau_{1}, \tau_{1}\right)\right)^{2}} \left[ \frac{\dot{\tilde{G}}\left(\tau_{1}, \tau_{1}\right)} {2\tilde{G}\left(\tau_{1}, \tau_{1}\right)} \left(x - \omega_{c}\right)^{\mu} - \dot{\omega}^{\mu}_{c} \right. \nonumber \\
	\left. - \frac{\left(x - \omega_{c}\right)^{\nu}} {2\tilde{G}\left(\tau_{1}, \tau_{1}\right)} \left(\delta^{\mu\nu} - \gamma^{\mu}\gamma^{\nu}\right) \right]\exp{\left(-\frac{\left(x - \omega_{c}\right)^{2}} {2 T \tilde{G}\left(\tau_{1}, \tau_{1}\right)}\right)}.
	\label{lowTf}
\end{align}
In the limit that $T$ vanishes we have a representation of the four dimensional $\delta-$function supported on the straight line joining the charges in Minkowski space. Applying Laplace's approximation allows us to determine the leading order behaviour which we illustrate for a point $x$ on the line from $a$ to $b$. In this case we clearly find the maximum at the point where $\omega_{c}\left(\tau_{0}\right) = x$ ($\tau_{0}$ remains unchanged from (\ref{tau0})) so the first and last terms vanish. Explicit application of (\ref{laplace}) results in
\begin{align}
	&\frac{q}{\left(2\pi T \tilde{G}\left(\tau_{0}, \tau_{0}\right)\right)^{\frac{3}{2}}} \frac{b - a}{\left|b - a\right|} \nonumber \\
	=&\frac{q}{\left(2\pi T\right)^{\frac{3}{2}}} \frac{\left|b - a\right|^{2}}{\left|x - a\right|^{\frac{3}{2}} \left|x - b\right|^{\frac{3}{2}}} \left(b - a\right)
\end{align}
which achieves a minimum at the midpoint of the line of magnitude
\begin{equation}
	\frac{q}{\left(\pi T\right)^{\frac{3}{2}}}.
\end{equation}
For sufficiently small $T$, this will be greater than the corresponding minimum associated to the bosonic theory.

This section contained some analytic results for bosonic and fermionic point particles beyond the leading order behaviour. It is especially interesting to note the small $T$ limit of the system which demonstrates a localisation of the field about the classical path between the charges. In the following section we return to the definition of the interaction between particles and use it to construct a full quantum theory.

\section{Contact interactions between particles}
\label{contact}
In this section we extend our work to describe a set of particles which interact when their worldlines intersect. Accordingly we work in a $D = 4$ dimensional space-time. This section carries out the analogous analysis to that in \cite{Us2} where a collection of strings interacting upon contact was considered. In this section we limit our discussion to bosonic particles. To describe the dynamics of a collection of interacting particles we augment the free action, $S_{0}$, of each point particle with a non-local contact interaction as follows:
\begin{equation}
	S_{\textrm{tot}} = \sum_{i} S_{0}\left[e_{i}, \omega_{i}\right] + \frac{g}{2}\sum_{ij} \int_{\omega_{i}} \!\int_{\omega_{j}} d\tau_{i}d\tau_{j}\,\dot{\omega}_{i}\left(\tau_{i}\right)  \cdot \dot{ \omega}_{j}\left(\tau_{j}\right) \delta^{4}\left(\omega_{i}\left(\tau_{i}\right) - \omega_{j}\left(\tau_{j}\right)\right)
	\label{sint}
\end{equation}
There are two ways to motivate this interaction. In correspondence with the introduction of \cite{Us2} one can substitute the expression for $\mathbf{E}^{\prime}$, the insertion used in Section \ref{Bosonic}, into the standard action of Maxwell electromagnetism for the energy of a static electric field:
\begin{equation}
	S_{E} = \frac{1}{4} \int d^{3} x\, \mathbf{E}\cdot\mathbf{E} = \frac{e^{2}}{4}\int_{\boldsymbol{\omega}}\!\int_{\boldsymbol{\omega}} d\tau_{1}d\tau_{2}\, \dot{\boldsymbol{\omega}}\left(\tau_{1}\right) \cdot \dot{\boldsymbol{\omega}}\left(\tau_{2}\right) \delta^{3}\left(\boldsymbol{\omega}\left(\tau_{1}\right) - \boldsymbol{\omega}\left(\tau_{2}\right)\right)
\end{equation} 
which produces a self-interaction along the worldline in question (we have integrated over the spatial point $x$). It is natural to then extend this to encompass interactions between distinct particles as we have done in (\ref{sint}). An alternative is to follow the worldline approach\footnote{I am grateful to the reviewer of this article for their insight in pointing this out.} and consider two particles which interact via the exchange of a single massive gauge boson between the points $\omega_{1}^{\mu}$ and $\omega_{2}^{\mu}$. The propagator for this process may be taken as $\Delta_{\mu\nu}\left(  \omega_{1}, \omega_{2}  \right) \propto \frac{e^{2}}{2}\int d^{4}p \frac{e^{  i p \cdot \left(   \omega_{1} - \omega_{2} \right)  }  } {  p^{2}+m^{2}  } $, which (as has been discussed by Feynman \cite{feyn} -- see also \cite{Schu}) should be inserted into the path integral and integrated with respect to the points of emission and absorption along each worldline. To take into account all possible interactions of this type, one is thus required to include a term
\begin{equation}
	\iint d\omega_{1}d\omega_{2}\, \Delta_{\mu\nu}\left(\omega_{1}, \omega_{2}\right)
\end{equation}
in the worldline theory. To achieve (\ref{sint}) from here it suffices to parameterise the paths by variables $\tau_{1}$ and $\tau_{2}$ and take the limit that the mediating boson has large mass so that $\frac{p^{2}}{m^{2}} \rightarrow 0$. This provides the appropriate $\delta$-function contact term, since the range of the interaction will be heavily supressed in this limit, and the resulting expression can be identified with (\ref{sint}) by setting $e^{2} = g m^{2}$.

The form of (\ref{sint}) may appear unusual but it is straightforward to verify that it satisfies the consistency criteria described by Kalb and Ramond for interactions of this type \cite{KalbR, KalbR2}. We also recall from the discussion in the introduction that the $\delta$-function ensures that the interaction is local in space-time. We must specify fixed boundary conditions for each particle, which we denote by $\omega_{i}\left(0\right) = a^{\mu}_{i}$ and $\omega_{i}\left(1\right) = b^{\mu}_{i}$. The coupling strength $g$ determines the relative strength of interactions between the worldlines (we have taken the couplings between all pairs of particles to be the same, rather than introducing separate coupling constants $g_{i}$ as presented in the introduction) and its sign determines whether the interaction is repulsive ($g > 0$) or attractive ($g < 0$). 

We shall consider the partition function of the theory described by $S_{\textrm{tot}}$ and determine its physical content as well as investigating whether there are divergences which need regularising. In order to build upon our earlier results which allowed us to relate the worldline theory to electrostatics we will retain the constraint on the path length of each worldline which was used in earlier sections, so that for each particle we impose $\int d\tau_{i}\, e_{i}\left(\tau_{i}\right) = T$. Since there is no reason to keep hold of this arbitrary dimensionful parameter we will follow the choice made in sections \ref{Bosonic} and \ref{Fermionic} by considering the large $T$ limit. We expect this limit to be the correct one for (\ref{sint}) to provide an extension of electrostatics and to ensure the interaction between particle worldlines is not unduly restricted to a localised region of space-time. In this theory we do not anticipate divergences corresponding to the coincidence of operators due to the well behaved nature of Green functions on one dimensional domains but we do stand to encounter unwanted divergences in taking the $T \rightarrow \infty$ limit. We will also find that the definition of the contact interaction will require slight refinement corresponding to the rather trivial vanishing of the argument of the $\delta$-function when $\tau = \tau^{\prime}$ and the worldlines $\omega_{i}$ and $\omega_{j}$ are the same.

Before proceeding we pause to point out how this action differs from the quantum mechanics which arises in both the action at a distance theories and the worldline formalism of a scalar quantum field. To compare with the action at a distance theories discussed in the introduction we may define a functional of the worldlines in order to recast the action above as if the particles interacted with a current. Let
\begin{equation}
	J^{\mu}\left(x\right) =  \sum_{i} \int_{\omega_{i}} d \omega_{i}^{\mu} \, \delta^{4} \left(x - \omega_{i}\right),
	\label{J}
\end{equation} 
which we know from section \ref{Bosonic} satisfies $\partial_{\mu}J^{\mu}\left(x\right) = \sum_{i} \delta^{4} \left(x - a_{i}\right) - \delta^{4}\left(x - b_{i}\right)$ so is sourced only at the points corresponding to the end of the worldlines (this is necessary in order that they be held in place). Then the interaction between the particles can be written as
\begin{equation}
	\frac{g}{2} \sum_{a} \int_{\omega_{a}} d\tau_{a}\, \,\dot{\omega}_{a}\left(\tau_{a}\right) \cdot J\left(\omega_{a}\left(\tau_{a}\right)\right)
	\label{Jint}
\end{equation}
and the equations of motion for the particles take the form (for illustration we gauge fix the einbein to $e = 1$ which imposes the first class energy-momentum constraint $\dot{\omega}^{2} = m^{2}$)
\begin{equation}
	m \frac{d}{d\tau}\left(\frac{\dot{\omega}_{a \mu}}{\sqrt{\dot{\omega}_{a}^{2}}}\right) = \frac{g}{2} \dot{\omega}_{a}^{\nu} \partial_{\left[\nu\right.}J_{\left. \mu \right]}
\end{equation}
which are well-known for the minimal vector coupling between worldlines and a background current\footnote{For completeness we point out that the equation for the ``field'' $J^{\mu}$ can be imposed by introducing a trivial space-time action $\frac{g}{4}\int d^{4}x \, J^{\mu}J_{\mu}$ whose variation, in conjunction with that of (\ref{Jint}), ensures the correct on-shell assignment in (\ref{J}). This is slightly different from the form of the theory discussed in the introduction where the interaction is truly non-local and is carried by a Green function with may be supported away from the null light cone. In that case the equations of motion and the space-time action would be familiar from minimal electromagnetism.}. We must recall, however, that this background current is not an independent degree of freedom because it depends on the particle worldlines (which must satisfy the equations of motion) and is also not a conserved quantity.

We also comment briefly on the relationship to the worldline formulation of quantum field theory to further justify its study. Although we have discussed how the interaction can be thought of as arising due to the exchange of very heavy gauge bosons between the worldlines, it can also be compared to the transition amplitude of the worldlines in some potential. If we consider a massless scalar field, $\phi$, with an interaction potential $U\left(\phi\right)$ (for simplicity we suppose that $\phi$ does not couple to any gauge field) then the one-loop effective action, $\Gamma\left[\phi\right]$, is defined by integrating over the quadratic quantum fluctuations of the matter degrees of freedom \cite{sBook}:
\begin{equation}
	\Gamma\left[\phi\right] = -\frac{1}{2}\textrm{Tr}\ln{\left(\frac{-\partial^{2} + U^{\prime \prime}\left(\phi\right)}{-\partial^{2}}\right)}.
\end{equation}
This is then rewritten in the worldline formalism by expressing the logarithm as an integral using the Schwinger proper time trick \cite{Schwinger} and the functional trace as a quantum mechanical transition amplitude for a point particle to traverse a closed loop:
\begin{align}
	\textrm{Tr}\ln{\left(\frac{-\partial^{2} + U^{\prime \prime}\left(\phi\right)}{-\partial^{2}}\right)} &\propto \int_{0}^{\infty}\! \frac{dT}{T} \!\int \!d^{D}\omega \left<\omega \big| \exp{\left[-T\left(-\partial^2 + m^2 + U^{\prime\prime}\left(\phi\left(\omega\left(\tau\right)\right)\right)\right]\right) } \big| \omega\right> \\ \nonumber
	& = \int_{0}^{\infty}\!\frac{dT}{T}\!\oint \mathscr{D}\omega \exp{\left(-\int_{0}^{1} d\tau \frac{\dot{\omega}^{2}}{2T} + m^2 T + U^{\prime\prime}\left(\phi\left(\omega\left(\tau\right)\right)\right)T\right)}
\end{align}
The effect of the self-interaction potential is to modify what would otherwise be the quantum theory of a free particle via an additional term $U^{\prime\prime}\left(\phi\left(\omega\left(\tau\right)\right)\right)$ in the action. In the case that $m = 0$ and there is no field potential the particle action reduces to $S_{0}$ which we used for functional averages in section \ref{Bosonic}. There we carried out the integral over $T$ by making use of the constraint on path lengths. 

The calculation of the effective action must now proceed perturbatively (for example, by expanding $\phi$ about the centre of mass of the worldline, which is the zero mode of the kinetic operator \cite{Auer}). However, if for the sake of illustration we wish to calculate the $2N$-point one-loop scattering amplitude, we may functionally differentiate $N$-times with respect to $\phi$ and then expand the field as a sum of $N$ plane waves:
\begin{equation}
	\phi\left(\omega\right) = \sum_{i = 1}^{N} e^{i p_{i} \cdot \omega}
\end{equation} 
Now, the physical picture of the interaction in (\ref{sint}) is of a pair of worldlines meeting at a point, where they interact before independently separating. This reminds us of the Feynmann diagram in $\phi^{4}$ theory so let us consider $U\left(\phi\right) = \frac{\lambda}{4!}\phi^{4}$, which leads to an additional worldline interaction $\frac{\lambda}{2}\phi\left(\omega\left(\tau\right)\right)^{2}$. Carrying out the Fourier integration over momentum the one-loop $2N$-point amplitude $\left<\phi\left(x_{1}\right)\ldots \phi\left(x_{2N}\right)\right>$ can be written in position space as a sum over permutations of pairings of the form
\begin{equation}
	\lambda^{N}\!\int_{0}^{\infty}dT\, T^{N - 1}e^{-m^{2}T}\int \!\mathscr{D}\omega\,\prod_{i = 1}^{N} \int_{0}^{1} d\tau_{i} \,\delta^{4}\left(\omega\left(\tau_{i}\right) - x_{2i}\right) \delta^{4}\left(\omega\left(\tau_{i}\right) - x_{2i - 1}\right)  e^{-\int_{0}^{1} d\tau \frac{\dot{\omega}^{2}}{2T}}.
\end{equation}
The interpretation of this is that the $\delta$-functions force the joining of pairs of points $x_{2i}$ and $x_{2i - 1}$ to a point on the virtual worldline $\omega\left(\tau_{i}\right)$, reproducing a quartic interaction vertex. The interaction that we propose in (\ref{sint}) is slightly different since we couple pairs of worldlines to one another directly without the need for a virtual loop to mediate the process but the interaction when pairs of lines intersect has the same form. The one-loop example considered here can be interpreted in terms of worldline interactions by imagining a particle travelling between two space-time points whose worldline intersects a second, closed worldline at a single point only.

Our integrand, however, does not consist only of the minimal scalar measures $\int d\tau_{i}$ but carries with it directional information\footnote{I wish to thank Paul Mansfield for helpful discussions on the physical interpretation of the interaction described in this paragraph.} through the factors of $\dot{\omega}^{\mu}$. So we imagine a four point contact interaction between pairs of worldlines which depends also on the tangent vectors to the worldlines at the point of contact. We get an idea of what this would mean for the corresponding field theory by dimensional analysis of the coupling strength. In $D$-dimensions, $\left[g\right] = \textrm{length}^{D - 2}$, whilst a scalar field has $\left[\phi\right] = \textrm{length}^{1 - \frac{D}{2}}$. If we introduce a coupling constant $\lambda$ then the field-theory interaction $L_{\textrm{int}} = \lambda \int \partial_{\mu} \phi \,\partial^{\mu} \phi \,\phi^{2}\, d^{D}x$ is dimensionless if the dimensions of $\lambda$ are the same as $g$. This suggests that the worldline theory proposed here is related to a field theory with quartic coupling containing two derivatives (we must recall, however, that the constraint on path lengths being imposed means that we are not reproducing the entire content of the field theory, which would require us to integrate over $T$ rather than consider only the contribution as $T \rightarrow \infty$). With this heuristic picture of how the contact interaction can be interpreted we now return to determine the partition function of the modified worldline theory.

We shall carry out the calculation as a perturbative expansion in $g$. Expressing the $\delta$-function in its Fourier representation introduces vertex operators
\begin{equation}
	\dot{\omega}\left(\tau_{i}\right) \cdot \dot{\omega}\left(\tau_{j}\right) \delta^{4}\left(\omega\left(\tau_{i}\right) - \omega\left(\tau_{j}\right)\right) = \int \frac{d^{4} k}{\left(2\pi\right)^{4}} V_{k}\left(\tau_{i}\right) \cdot V_{-k}\left(\tau_{j}\right)
\end{equation}
At first order in $g$ the correction to the partition function of the non-interacting theory takes the form
\begin{equation}
	\frac{g}{2} \sum_{j k} \int \left(\prod_{i} \frac{\mathscr{D} \left(\omega_{i}, e_{i}\right)}{\mathcal{Z}}\delta\left(\int e_{i} d\tau_{i} - T\right)e^{-S_{0}\left[e_{i}, \omega_{i}\right]} \right)\int\!\!\int d\omega_{j} \cdot d\omega_{k} ~\delta^{4}\left(\omega_{j} - \omega_{k}\right)
	\label{og}
\end{equation}
There are two contributions to this sum. When the worldlines are distinct (\ref{og}) can be factorised to make use of the result in section \ref{Bosonic}:
\begin{align}
	&\frac{g}{2}\sum_{j \neq k} \int \left(\prod_{i \neq j,k}\frac{\mathscr{D} \left(\omega_{i}, e_{i}\right)}{\mathcal{Z}}\delta\left(\int e_{i} d\tau_{i} - T\right)e^{-S_{0}\left[e_{i}, \omega_{i}\right]} \right) \times \nonumber \\
	 &\qquad\int \frac{\mathscr{D} \left(\omega_{j}, e_{j}\right)}{\mathcal{Z}}\delta\left(\int e_{j} d\tau_{j} - T\right)e^{-S_{0}\left[e_{j}, \omega_{j}\right]} \int d\omega_{j}^{\mu} \left<\int d\omega_{k}^{\mu}\delta^{4} \left(\omega_{j} - \omega_{k}\right)\right>_{T}= \\
	 & \frac{g}{8\pi^{2}}\! \sum_{j \neq k} \int \frac{\mathscr{D} \left(\omega_{j}, e_{j}\right)}{\mathcal{Z}}\delta\left(\int e_{j} d\tau_{j} \!- \!T\right)e^{-S_{0}\left[e_{j}, \omega_{j}\right]} \int d\omega_{j}^{\mu} \frac{\partial}{\partial \omega_{j}^{\mu}} \left(\frac{1}{\left|\omega_{j} - a_{k}\right|^{2}} - \frac{1}{\left|\omega_{j} \!-\! b_{k}\right|^{2}}\right)
\end{align}
where we have taken the large $T$ limit and discarded contributions of order $\frac{1}{T}$ arising from the expectation value in the first line.
The integral over $\omega_{j}$ at the far right of the bottom line produces a boundary contribution and we recall that the boundary conditions ensure this is a constant throughout the functional integral over $\omega_{j}$. So the functional integral is rendered trivial and the result is
\begin{equation}
	\frac{g}{4\pi^{2}} \sum_{j \neq k} \left[\left(\frac{1}{\left|a_{j} - a_{k}\right|^{2}} - \frac{1}{\left|a_{j} - b_{k}\right|^{2}}\right) - \left(\frac{1}{\left|b_{j} - a_{k}\right|^{2}} - \frac{1}{\left|b_{j} - b_{k}\right|^{2}}\right)\right].
	\label{distinct}
\end{equation}
The interpretation of this is a pairwise interaction between the fixed ends of the open worldlines of the theory, whose form we recognise as the space-time Green function $\left(\partial^{2}\right)^{-1}$. This is the Green function of the scalar field theory whose worldline reformulation leads to the quantum theory described by $S_{0}$ which we have discussed above. The effect of the contact interaction is to couple the end points of the worldlines to one another via the propagators in (\ref{distinct}). This retrospectively suggests that the high $T$ limit was a good choice, since it leads to a familiar interaction with the long-distance properties that one would hope for in quantum field theory, in much the same way that the same limits taken in sections \ref{Bosonic} and \ref{Fermionic} yielded familiar electrostatics fields.

This can be contrasted with the result we found in the case of interacting strings in \cite{Us1} and \cite{Us2}. We have already explained how the endpoints of the strings were fixed to lie on some specified worldlines. The effect of the contact interaction on the worldsheet came primarily from the region close to its boundary, $B$, and provided an interaction between the worldlines which took the form
\begin{equation}
	\sum_{a, b} \frac{1}{2\pi^{2}} \int_{B_{a}}\!\int_{B_{b}} \frac{d\omega_{a} \cdot d\omega_{b}}{\left|\omega_{a} - \omega_{b}\right|^{2}}\,,
	\label{resString}
\end{equation}
which now contains the space-time propagator of a vector field integrated over all points on the boundary of each string. In fact the Fourier transform of the Green function took the form $\delta_{\mu\nu}\left(k\right) = \eta_{\mu\nu} / k^{2} - k_{\mu}k_{\nu} / (k^{2})^{2}$, which is the propagator for an Abelian gauge boson, $\mathcal{A}$ in the gauge $\partial \cdot \mathcal{A} = 0$. In this way we related the tensionless limit of the interacting string theory to quantum electrodynamics, where the contact interaction served to produce the interactions between the worldlines and the gauge field. The calculation of the partition function of the point particle theory is related to (\ref{resString}) as analogous to the contribution arising from two fixed points in space-time. If these are taken to be points $\omega_{a}$ and $\omega_{b}$ on two curves $B_{a}$ and $B_{b}$ then the string theory result arises by then integrating over all such points on the curves (see also the end of section \ref{Bosonic} where we discussed the transition from a spatial average to averaging over curves in space-time). In the string theory version this integral is generated by the extra structure in the worldsheet contact interaction which produces the integration measure and automatically counts the contributions from every point on the string boundary.

Returning now to the point particle theory we must address the case $j = k$ separately as then the worldlines being integrated over are the same. In the string theory case \cite{Us1, Us2} we found that two contributions make up this interaction -- a renormalisation of the free string action and cosmological constant and a contribution corresponding to self-intersection of the string. For point particles, however, no such self-intersection ought to be present since the vanishing of 
\begin{equation}
	\frac{g}{2}\int\!\!\int d\tau_{1} d\tau_{2} ~\dot{\omega}\left(\tau_{1}\right) \cdot \dot{\omega}\left(\tau_{2}\right) ~ \delta^{4}\left(\omega\left(\tau_{1}\right) - \omega\left(\tau_{2}\right)\right)
\end{equation}
is only at $\tau_{1} = \tau_{2}$. Naively this would provide $\frac{g}{2}\delta\left(0\right) \times \textrm{Length}\left(\omega\right)$, suggestive of a renormalisation of the non-interacting part of the action. A second major difference between the theory of point particles we consider here and the string theory equivalent is that the current case does not require us to worry about encountering a conformal anomaly. In particular there are no short distance divergences associated to the Green function which would require regularisation. To make all this more precise we must consider
\begin{equation}
	\left<I^{\mu\nu}\right>_{T} = \int \frac{d^{4} k}{\left(2\pi\right)^{4}}\int\!\!\int d\tau_{1} d\tau_{2}\, \frac{g}{2}\left< \frac{d \omega^{\mu}}{d\tau_{1}}  e^{i k \cdot \omega\left(\tau_{1}\right)} e^{-i k \cdot \omega\left(\tau_{2}\right)}\frac{d \omega^{\nu}}{d\tau_{2}}\right>_{T}
\end{equation}
where we have again used the Fourier representation of the four-dimensional $\delta$-function. The insertions are once more easily generated via the introduction of sources and the integral over $\omega$ provides the following generalisation of (\ref{integrated})
\begin{align}
	\frac{g}{2}&\int \frac{d^{4} k} {\left(2\pi\right)^{4}}\int\!\!\int d\tau_{1} d\tau_{2}\,\frac{\delta}{\delta j^{\mu}\left(\tau_{1}\right)} \frac{\delta}{\delta j^{\nu}\left(\tau_{2}\right)} \nonumber \\
	&\exp{\left(-\frac{1}{2} \sum_{i,j = 1}^{2} k_{i}\cdot k_{j} \,G\left(\tau_{i}, \tau_{j}\right) - i \int d\tau \,\sum_{i = 1}^{2} k_{i} \cdot j\left(\tau\right) \frac{d}{d \tau}G\left(\tau_{i}, \tau\right) \right.} \nonumber \\
	&\left. \hphantom{\exp{\left(-\frac{1}{2} \sum_{i,j = 1}^{2} k_{i}k_{j} G\left(\tau_{i}, \tau_{j}\right)\right)}} + \frac{1}{2} \int\!\int d\tau d\tau^{\prime}  \,j\left(\tau\right) \cdot j\left(\tau^{\prime}\right) \frac{d}{d\tau} \frac{d}{d\tau^{\prime}} G\left(\tau, \tau^{\prime} \right) \right) \nonumber \\
	&\left.\hphantom{\exp{\left(-\frac{1}{2} \sum_{i,j = 1}^{2} k_{i}k_{j} G\left(\tau_{i}, \tau_{j}\right)\right)}}\exp{\left( - \int d\tau\, j\left(\tau\right) \cdot \frac{d}{d\tau} \omega_{c}\left(\tau\right) + \sum_{i = 1}^{2} i k_{i} \cdot \omega_{c}\left(\tau_{i}\right) \right)} \right|_{j = 0}
\end{align}
where $k_{1} = k = -k_{2}$. It is trivial to carry out the functional differentiation and it proves useful to define 
\begin{equation}
\Psi\left(\tau_{1}, \tau_{2}\right) \equiv G\left(\tau_{1},\tau_{1}\right) + G\left(\tau_{2},\tau_{2}\right) - 2G\left(\tau_{1}, \tau_{2}\right) 
\end{equation}
in order to express the answer as
\begin{align}
	&\frac{g}{2}\int_{0}^{1}d\tau_{1}\int_{0}^{1}d\tau_{2}~\exp{\left(-\frac{1}{2}k^{2}\Psi\left(\tau_{1}, \tau_{2}\right)\right)} \left[\dot{\omega}_{c}^{\mu}\dot{\omega}_{c}^{\nu} - \frac{i}{2}k^{\mu} \dot{\omega}_{c}^{\nu}d_{t}^{2}\Psi\left(\tau_{1}, \tau_{2}\right)  \right. \nonumber \\
	&\qquad \qquad \qquad\left. + \frac{i}{2} k^{\nu} \dot{\omega}^{\mu} d_{t}^{1}\Psi\left(\tau_{1}, \tau_{2}\right) - \frac{1}{2}\left(\eta^{\mu\nu} - \frac{k^{\mu}k^{\nu}}{k^{2}}\right) d_{t}^{1}d_{t}^{2} \Psi\left(\tau_{1}, \tau_{2}\right) \right] e^{i k \cdot \left( \omega_{c}\left(\tau_{1}\right) - \omega_{c}\left(\tau_{2}\right)\right)}
	\label{twoints}
\end{align}
which must then be integrated over $k$. In the above equation we have integrated by parts to produce the transverse projector for the final term in square brackets. In the worldline formalism it is often advantageous to follow an alternative integration by parts procedure advocated by Bern and Kosower \cite{BK3, Bern} which removes all second derivatives of Green functions\footnote{The worldline approach was originally proposed by Strassler as a means of arriving at earlier results relating field theory amplitudes to the infinite tension limit of string theory \cite{BK1, BK2}, so that many calculational tricks used by string theorists are borrowed for worldline calculations.}. The interpretation of this technique has its roots in the pinching of Feynman diagrams present in the underlying field theory, but as we do not necessarily have such a model behind our work it is unnecessary for us to adhere to it. Now 
\begin{equation}
	\Psi\left(\tau_{1}, \tau_{2}\right) = T\left(\left|\tau_{1} - \tau_{2}\right| - \left(\tau_{1} - \tau_{2}\right)^{2}\right)
	\label{PsiG}
\end{equation}
actually coincides with twice the Green function with periodic boundary conditions \cite{Dai1}. As such it satisfies $\frac{-1}{2T}\frac{d^{2}}{d\tau_{1}^{2}}\Psi\left(\tau_{1}, \tau_{2}\right) = \delta\left(\tau_{1} - \tau_{2}\right) - 1$ and is a function of $\tau_{1} - \tau_{2}$ only. Noting also that $\omega_{c}\left(\tau_{1}\right) - \omega_{c}\left(\tau_{2}\right) = \left(b - a\right) \left(\tau_{1} - \tau_{2}\right)$ we learn that the entire expression (\ref{twoints}) is in fact a function of the separation $\tau_{1} - \tau_{2}$. We use this to fix the zero by setting $\tau_{2} = 0$ and multiplying by $\int_{0}^{1}d\tau_{2} = 1$. Furthermore $\Psi\left(\tau_{1}, 0\right) = T\tau_{1}\left(1 - \tau_{1}\right)$ is just the coincident Green function we met in section \ref{Bosonic} so we must determine
\begin{align}
	g\int_{0}^{1}d\tau_{1} \exp{\left(-\frac{1}{2}k^{2}G\left(\tau_{1}, \tau_{1}\right)\right)} \left[\dot{\omega}_{c}^{\mu}\dot{\omega}_{c}^{\nu} +\frac{i}{2}k^{\mu} \dot{\omega}_{c}^{\nu}d_{t}^{1}G\left(\tau_{1}, \tau_{1}\right) + \frac{i}{2} k^{\nu} \dot{\omega}^{\mu} d_{t}^{1}G\left(\tau_{1}, \tau_{1}\right)\right. \nonumber \\
	\left. + \frac{1}{2}\left(\eta^{\mu\nu} - \frac{k^{\mu}k^{\nu}}{k^{2}}\right) d_{t}^{1}d_{t}^{1} \Psi\left(\tau_{1}, 0\right)\right] e^{i k \cdot \left(b - a\right) \tau_{1}}.
	\label{renorm}
\end{align}
Anticipating that we will eventually take the limit $T \rightarrow \infty$ we need only consider an expansion in powers of $\frac{1}{T}$. The leading order contributions again come only from the ends of the interval where the coincident Green function vanishes so it suffices to expand about $\tau_{1} = 0$ and $\tau_{1} = 1$.  We shall consider each term in (\ref{renorm}) separately; at lowest order in $\frac{1}{T}$ the first takes the form
\begin{align}
&g \dot{\omega}_{c}^{\mu} \dot{\omega}_{c}^{\nu}  \int d\tau_{1} \,e^{-\frac{1}{2}k^{2}G\left(\tau_{1},\tau_{1}\right)} e^{i k \cdot \left(b - a\right)\tau_{1}}  \nonumber \\
=\,& \frac{2g} {k^{2}}\left[e^{i k \cdot \left(b - a\right)} - 1\right] \frac{\dot{\omega}_{c}^{\mu} \dot{\omega}_{c}^{\nu}} {T} + \mathcal{O}\left(\frac{1}{k^{4}T^{2}}\right)
\end{align}
Upon integrating over $k$ the latter term in square brackets vanishes whilst the former provides $\left(2\pi\left|b - a\right|\right)^{-2}$. The contribution we seek is the trace of this -- we note that since the derivative of the classical solution to the equations of motion is a constant we may express the above expression as 
\begin{equation}
	\frac{g}{\pi^{2}\left|b - a\right|^{2}} \int_{0}^{1} d\tau \, \frac{\dot{\omega}_{c}^{2}}{2T} + \mathcal{O}\left(\frac{1}{k^{4}T^{2}}\right).
\end{equation}
We interpret this as providing a renormalisation of the free action and note that it is suppressed in the large $T$ limit. 

The second and third terms in (\ref{renorm}) involve a derivative of the Green function and as such provide contributions that are independent of $T$. Taking the trace we require
\begin{align}
	&gi k \cdot \dot{\omega}_{c} \int d\tau_{1} \, \dot{G}\left(\tau_{1} \tau_{1}\right) e^{-\frac{1}{2}k^{2}G\left(\tau_{1},\tau_{1}\right)} e^{i k \cdot \left(b - a\right)\tau_{1}} \nonumber \\
	=\,&g \frac{i k \cdot \dot{\omega}_{c} }{2 k^{2}} \left[e^{i k \cdot \left(b - a\right)} - 1\right]  + \mathcal{O}\left(\frac{1}{k^{2}T}\right)
\end{align}
Carrying out the integral over $k$ provides
\begin{equation}
	-g \frac{\dot{\omega}_{c} \cdot \left(b - a\right)}{4\pi^{2} \left|b - a\right|^{4}} + \mathcal{O}\left(\frac{1}{k^{2}T}\right)
\end{equation}
Recognising that $\dot{\omega} = b - a$ we may also cast this into the form
\begin{equation}
	-\frac{g}{4\pi^{2} \left|b - a\right|^{2}} \int_{0}^{1} \frac{\dot{\omega}_{c}^{2}}{\left|b - a\right|^{2}} + \mathcal{O}\left(\frac{1}{k^{2}T}\right)
\end{equation}
providing a finite renormalisation of the free action.

Finally we take the last term of (\ref{renorm}) and consider its trace. We use the defining equation of the Green function to write its contribution as
\begin{equation}
	-g \int_{0}^{1} d\tau_{1} \, T\left(\delta\left(\tau_{1}\right) - 1\right) e^{-\frac{1}{2}k^{2}G\left(\tau_{1},\tau_{1}\right)} e^{i k \cdot \left(b - a\right)\tau_{1}}
	\label{Delta}
\end{equation}
Both terms are independent of the field $\omega_{c}$ which suggests that we should interpret them as renormalisations of a cosmological constant term in the action $\int_{0}^{1} d\tau \,T$. The $\delta$-function gives us
\begin{equation}
	-g \int_{0}^{1}d\tau ~ T
\end{equation}
exactly as required, albeit formally divergent when we take $T$ to infinity\footnote{The integral over $k$ also provides an infinite volume factor multiplying this result. We may tidy this up by carrying out the $k-$integral before integrating over $\tau_{1}$. Doing so turns (\ref{Delta}) into 
\begin{equation}
	-g \int d\tau_{1} \, \left( \frac{\pi} {2G\left(\tau_{1}, \tau_{1}\right)} \right)^{\frac{D}{2}} e^{-\frac{\left(b - a\right)^{2}\tau_{1}^{2}} {2G\left(\tau_{1}, \tau_{1}\right)} }  \left(\delta\left(\tau_{1}\right) - 1\right).
\end{equation} 
We are concerned only with the contribution arising from the $\delta$-function so we need the value of the exponent as $\tau_{1} \rightarrow 0^{+}$. An easy calculation shows that the exponent vanishes leaving only 
\begin{equation}
	-g\left(\frac{\pi}{2}\right)^{\frac{D}{2}} \lim_{\tau_{1} \rightarrow 0} \left(  G\left(\tau_{1}, \tau_{1}\right)  \right)^{-\frac{D}{2}}
\end{equation}
which diverges.}, whereas the second evaluates to
\begin{equation}
	-\frac{2g}{ k^{2}} \left[e^{i k \cdot \left(b - a\right)} - 1\right] + \mathcal{O}\left(\frac{1}{k^{2}T}\right),
	\label{gk}
\end{equation}
which upon integrating over $k$ becomes
\begin{equation}
	-\frac{g}{2\pi^{2}}\frac{1}{\left|b - a\right|^{2}} + \mathcal{O}\left(\frac{1}{k^{2}T}\right).
	\label{int1}
\end{equation}
This surprisingly takes the same form as (\ref{distinct}) and suggests that there is after all a self-interaction present in the theory, sensitive only to the boundary of the worldline. 

To lend this more weight we could approach the calculation in a complementary fashion. We note that $\Psi\left(\tau_{1}, \tau_{2}\right)$ vanishes only at $\tau_{2} = \tau_{1}$ so we could arrange our calculation by instead expanding about this point. When $\tau_{1}$ is not close to the boundary (measured with respect to $\frac{1}{k^{2}T}$) integrating $\tau_{2}$ about $\tau_{1}$ corresponds to integrating over their relative separation and the leading order contribution arises by setting the $T$-independent exponent $\exp{\left(i k \cdot \left(b - a\right)\left(\tau_{1} - \tau_{2}\right)\right)}$ equal to unity. This gives the terms above which are absent of an exponent. When $\tau_{1}$ is close to the boundary we must take care because $\tau_{2}$ is restricted to lie in $\left[0, 1\right]$. So for example when $\tau_{1} \approx 0$ we must integrate $\tau_{2}$ a small distance from this boundary along the line but must also consider the contribution when $\tau_{2}$ is integrated from the opposite boundary along the line. Indeed for the latter case we consider, following the notation of section \ref{Bosonic}, 
\begin{equation}
	\frac{g}{2}\int_{0}^{h}d\tau_{1} \int_{1-h}^{1}d\tau_{2}~ T\left(\delta\left(\tau_{1} - \tau_{2}\right) - 1\right) e^{-\frac{1}{2} k^{2} \Psi\left(\tau_{1}, \tau_{2}\right)} e^{i k \cdot\left(b - a\right)\left(\tau_{1} - \tau_{2}\right)}.
\end{equation}
The $\delta$-function is not supported for this configuration of variables, besides it is the second term with which we are concerned. At leading order in $\frac{1}{T}$ we evaluate the trailing exponent at the point $\tau_{1} = 0$, $\tau_{2} = 1$. A change of variables $\tau = 1 - \tau_{2}$ yields $\Psi\left(\tau_{1}, 1 - \tau\right) = T\left(\left(\tau_{1} + \tau\right) - \left(\tau_{1} + \tau\right)^{2}\right)$ and the integral becomes
\begin{equation}
	-\frac{g}{2} \int_{0}^{h} d\tau_{1} \int_{0}^{2h} du~ e^{-\frac{1}{2}k^{2}G\left(u, u\right)} e^{-i k \cdot\left(b - a\right)}
\end{equation}
where we have set $u = \tau_{1} + \tau$. At leading order this provides
\begin{equation}
	-\frac{g}{k^{2}}e^{-i k \cdot\left(b - a\right)}
\end{equation}
which is to be compared with (\ref{gk}). The sign of the exponent is not important and the factor of two is found by including the configuration where the positions of $\tau_{1}$ and $\tau_{2}$ in the above calculation are swapped. So the physical information arises when the two points are close to opposite boundaries, whereas the renormalisations appear from the region where the two points coincide. This is in fitting with the naive analysis of the form of the contact interaction when the worldlines are the same.

The self-interaction (\ref{int1}) also has the appropriate factor of two difference for it to be subsumed into the sum (\ref{distinct}) so that at first order in the expansion of the interaction we find
\begin{equation}
	\frac{g}{4\pi^{2}} {\sum_{j, k}}^{\prime} \left[\left(\frac{1}{\left|a_{j} - a_{k}\right|^{2}} - \frac{1}{\left|a_{j} - b_{k}\right|^{2}}\right) - \left(\frac{1}{\left|b_{j} - a_{k}\right|^{2}} - \frac{1}{\left|b_{j} - b_{k}\right|^{2}}\right)\right].
	\label{all}
\end{equation}
We write $\sum^{\prime}$ to denote that when $j = k$ we discard the first and last terms in the summand where the separation vanishes. This concludes the analysis of the first order effect of the contact interaction present in this theory. In the following section we turn to consider higher order interactions and show that the result above is not spoilt and that this propagator interaction between the end of the worldlines exponentiates.

\subsection{Expansion to arbitrary order}
We may consider expanding the part of the action corresponding to the inter-particle interaction to arbitrary order in the coupling strength. When the interaction is between distinct worldlines we may repeat the analysis at first order to simplify the calculation to a form familiar from section \ref{Bosonic}. Here we address the problem of having a fixed even number, $N$, of points to be integrated over the same worldline which could potentially spoil the result at order $g^{\frac{N}{2}}$ or higher. The reason for this is that we must be cautious in taking the limit $T \rightarrow \infty$, in case the clustering of points on the worldline produces dependence on $T$ which would cause the result to diverge. So we consider
\begin{equation}
	\left<V^{\mu}_{k_{1}}\left(\tau_{1}\right)V^{\nu}_{k_{2}}\left(\tau_{2}\right) \ldots V^{\alpha}_{k_{N}}\left(\tau_{N}\right)\right>_{T}
\end{equation}
where we understand that each point must be integrated from $\tau_{i} = 0$ to $1$. 

Wick's theorem produces a common factor to all of the contractions that could be formed out of the above product of fields which takes the form
\begin{equation}
	\exp{\left(-\frac{1}{2} \sum_{i,j = 1}^{N} k_{i} \cdot k_{j} G\left(\tau_{i}, \tau_{j}\right)\right)} e^{i \sum_{i}k_{i} \cdot \omega\left(\tau_{i}\right)}
	\label{WickExp}
\end{equation}
We are interested in taking the high $T$ limit, whereby the leading contribution will be from the regions of integration where the exponent vanishes. There are no short distance divergences which require regularisation so we may consider the points to be arbitrarily close to one another. This leads us to consider two cases. The exponent can be made to vanish by ensuring that each individual Green function $G\left(\tau_{i}, \tau_{j}\right)$ vanishes, which requires us to localise each of the points close to either end of the domain. Alternatively we might consider bringing a collection of points coincident, which is instructive in order to compare with the results of the previous subsection. As no regularisation is required we keep the discussion of this case brief. It is useful to split (\ref{WickExp}) as 
\begin{equation}
	\exp{\left(-\frac{1}{2}  \sum_{i = 1}^{N} k_{i}^{2} G\left(\tau_{i}, \tau_{i}\right) - \sum_{j \neq i} k_{i} \cdot k_{j} G\left(\tau_{i}, \tau_{j}\right)  \right)} e^{i \sum_{i}k_{i} \cdot \omega\left(\tau_{i}\right)}
\end{equation}
which makes it clear that for an arbitrary configuration of the points away from the boundary the first factor damps the integrand. Now suppose that some number, $n$, of these points are brought to the same point $\tau_{1}$. Then these points contribute
\begin{equation}
	\exp{\left(-\frac{1}{2} G\left(\tau_{1}, \tau_{1}\right)  \left( \sum_{i = 1}^{n} k_{i}\right)^{2} + \ldots \right) } e^{i \sum_{i}k_{i} \cdot \omega\left(\tau_{i}\right)}
\end{equation}
where the $\ldots$ represents terms which depend on the relative separation between each point and $\tau_{1}$. Recall that $G\left(\tau_{1}, \tau_{1}\right) = T\tilde{G}\left(\tau_{1}, \tau_{1}\right)$ vanishes only at the boundary of the domain. Consider how this exponent behaves under the integral with respect to $k_{1}$, say. In the large $T$ limit a Laplace approximation implies its effect is to provide\footnote{This should be considered in light of the results at first order where we had $k_{1} = k = -k_{2}$ where the exponent vanished throughout the domain when $\tau_{1} = \tau_{2}$. We exclude such cases here because we understand that they lead to a simple renormalisation of the action.}
\begin{equation}
	\frac{ \delta\left(\sum_{i}k_{i}\right) }{\left(2T \tilde{G}\left(\tau_{1}, \tau_{1}\right)\right)^{2}}.
	\label{delta}
\end{equation}
Since in one dimension the coincident Green function is finite we see that all contributions from such a configuration are suppressed by a factor of $T^{-2}$. The size of the integrals over the relative separation $\tau_{i} - \tau_{1}$ can be examined by power counting in $\frac{1}{T}$. The largest contribution arises when all fields take part in a contraction and can be arranged into a series of $\frac{n}{2}$ second order derivatives. A simple calculation shows that this contribution is of order $T^{\frac{n}{2}}$. But after integrating the other $n - 1$ points about $\tau_{1}$ the resulting expression is of order $T$. In combination with (\ref{delta}) this contribution can be seen to be subleading in $\frac{1}{T}$, so no unwanted divergences are encountered. 

Returning to the case that all of the points are on the boundary, each of the Green functions in (\ref{WickExp}) vanishes. So we expect a contribution to the expectation value from the region of integration where each point is close to the boundary. We have learnt in the previous subsection that the $\mathcal{O}\left(1\right)$ term arises when we contract the fields inside each vertex operator amongst themselves. The exponents $\exp{\left( i k \cdot \omega_{c}\left(\tau_{i}\right)\right)}$ can be approximated at leading order by replacing them with their values at the appropriate boundary -- we denote this by $\exp{\left( i k \cdot \omega_{B_{i}}\right)}$. So we consider a term with $r$ such contractions:
\begin{align}
&	\prod_{j = 1}^{r}\left(\int_{0}^{h}\! + \!\int_{1 - h}^{h}\right)d\tau_{j} ~i k_{j}^{\mu_{j}} \,\dot{G}_{jj} e^{-\frac{1}{2} k^{2} G_{jj}}e^{i k_{j} \cdot \omega_{B_{j}}}\! \prod_{i = r + 1}^{N} \left(\int_{0}^{h} \!+\! \int_{1 - h}^{h}\right)d\tau_{i}~ \dot{\omega}_{ci}  e^{-\frac{1}{2} k^{2} G_{ii}}e^{i k_{i} \cdot \omega_{B_{i}}} \nonumber \\
	=& \prod_{j = 1}^{r} \frac{i k_{j}^{\mu_{j}}}{k_{j}^{2}}\left(e^{i k_{j} \cdot a} - e^{i k_{j} \cdot b}\right)\prod_{i = r + 1}^{N} \left[\frac{\dot{\omega}_{ci}}{k_{i}^{2} T} + \mathcal{O}\left(\frac{1}{\left(k^{2} T\right)^{2}}\right)\right]
\end{align}
The leading order contribution clearly requires $r = N$ so that all fields are contracted, whereby the above expression reduces to
\begin{equation}
	\prod_{j = 1}^{N} \frac{i k_{j}^{\mu_{j}}}{k_{j}^{2}}\left(e^{i k_{j} \cdot a} - e^{i k_{j} \cdot b}\right).
\end{equation}
We now impose pairwise $k_{i + 1} = -k_{i}$ and contract $\mu_{i}$ and $\mu_{i+1}$ to produce the expectation value of the contact interaction at order $g^{\frac{N}{2}}$:
\begin{equation}
	-2^{\frac{N}{2}} g^{\frac{N}{2}} \prod_{i = 1}^{\frac{N}{2}} \frac{e^{i k \cdot \left(b - a\right)}}{k_{i}^{2}}.
\end{equation}
As our analysis has been for an arbitrary number of points the result at this order is not spoilt by considering a higher order term in the expansion of the contact interaction. The above equation is simply the Fourier space version of the order $g^{\frac{N}{2}}$ contribution to the exponential of the sum of boundary interactions (\ref{all}), from which we conclude that this interaction between worldlines exponentiates absent of short-distance divergences. 

\section{Discussion}
In this paper we have considered contact interactions in the context of theories of point particles. We have demonstrated a novel way of generating the static electric field for a pair of point particles by considering fluctuating worldlines whose endpoints are fixed to the positions of the charges. The functional approach we used for calculation allowed us to generalise this construction to include spin 1/2 particles and we then used the formalism to construct a quantum theory describing a set of point particles interacting upon contact. The result for fermionic particles is interesting because it suggests an unusual electric force acting on Dirac spinors due to the charged particles. Both results depend somewhat on the choice of boundary conditions imposed on the worldline fields at either end of the paths and we also found it necessary to introduce constraints on the worldline metric and its super-partner to reconstruct the classical field of static charged particles.

We demonstrated that the worldline contact interaction provides an unconventional method of generating the static dipole electric field due to equal and oppositely charged particles, and also modified the boundary conditions on the curves to produce the field of a static point charge. This result required us to take the intrinsic lengths of the curves, $T$, to be very large but we also investigated the corrections to the fields for finite values of $T$. The inverse square law for electrostatic interactions has been verified to a very high precision which provides a bound on the size of the deviations from this behaviour. This in turn allows us to find a minimum bound for $T$. An alternative method of estimating the size of $T$ follows from the $1 / T$ corrections to the retarded Green function given in equation (\ref{speed}) which are discussed in section \ref{FiniteT}. This would imply corrections to the speed of propagation of light, $c$, which is also known to extremely high precision. 

Should future experiment further refine these two quantities then we may increase our lower bound on $T$, or we may test unexpected deviations against the results provided by the model in this paper. Either way, from a theoretical standpoint varying the parameter $T$ has an important effect on the form of the field lines, especially when we consider the low $T$ behaviour where the novel behaviour of confining field lines was discovered. It is important to better understand this limiting case because it may help to develop another approach to this well known aspect of non-Abelian gauge field theories. The worldline techniques we have used here may prove valuable in approaching this problem because of the computational simplicity present with this approach.

Beyond the first order calculations we presented in the first few sections of this paper we also determined the partition function for a set of particles which interact when their worldlines intersect. We showed how the contact interaction produces propagator couplings between the fixed endpoints of the worldlines, which is not unexpected considering that the effect of the interaction is to modify the underlying scalar field theory to include a quartic self-coupling. An important avenue for further investigation would be to work out the exact relationship between the quartic field theory and the interacting worldlines, since it may then prove possible to rephrase old questions in field theory in terms of simple quantum mechanics on the worldlines, adding a calculational tool which readily inherits the efficiencies present within the worldline formalism of quantum field theory.

The worldline formalism highlights an intimate connection between field theory and the first quantised particles we have considered here. Our approach differed in that rather than coupling the theory living on the worldline to a background gauge field (as occurs when integrating out matter fields with the worldline approach) we coupled the worldlines to one another directly using the $\delta-$function interaction. We also introduced fundamental constraints on the worldlines' lengths (rather than integrating over $T$) and worked only with open worldlines. However, the functional methods that we used are not limited to the calculations we have undertaken in this instance and can be directly applied to other situations that arise in the worldline approach. 

It would be interesting to consider a similar contact interaction defined on \textit{closed} worldlines since this case is encountered in the determination of one-loop effective actions and multi-loop scattering amplitudes in the worldline approach. There are substantial differences between the closed worldlines and the open paths that we have considered in this paper, in part due to the difference in their topology. In particular the Green function is related to $\Psi(\tau_{1}, \tau_{2})$ of (\ref{PsiG}), the boundary conditions requiring it to be periodic rather than vanishing at the ends of the interval. In this case, the Green function's coincident limit is constant along the worldline (momentum conservation implies that, in flat space, the numerical value of the coincident limit is not important). In this article we relied on the functional form of the Green function to ensure that in the high $T$ limit our integrands were suppressed moving away from the endpoints of the worldline which would no longer hold true if the particle's path were closed. It would thus be necessary to improve the analytical calculation of the integrals instead of extracting the leading order behaviour as we have in previous sections. At higher loop orders the worldlines have extra insertions of propagators and so we would like to generalise this work to the case of an arbitrary one dimensional topology (the Green functions for such worldlines have been studied extensively in \cite{Dai1, Dai2}). This would signify useful progress in the first quantised formalism.

We have also discussed how the calculations presented here for worldlines with fixed endpoints are related to an analogous theory of interacting strings, where the worldline formalism allowed us to relate the string theory to products of Wilson lines in spinor quantum electrodynamics. As in this paper, it was also necessary to take a particular limit in order to secure that result, namely the limit of vanishing string tension. This has the effect of making the strings macroscopically large in comparison to the worldlines (which are their boundaries), just as the $T \rightarrow \infty$ limit discussed here makes the worldlines' lengths very large. In contrast to this paper the tensionless limit of the string theory was necessary to suppress divergences which would otherwise spoil the conformal invariance of the classical theory, which presents a fundamental obstruction to investigating finite tension corrections. In this sense, the worldline theory has an advantage over the interacting string theory, in that we may vary $T$ without encountering any quantum anomalies. 

\subsubsection*{Acknowledgements}
The author takes great pleasure in thanking Paul Mansfield, David Berman and Paul Heslop for useful discussions and in acknowledging valuable comments on the manuscript from Daniele Galloni, Thomai Tsiftsi and Paul Mansfield. The research presented here was supported by STFC via a studentship and in part by the Marie Curie network GATIS (gatis.desy.eu) of the European Union's Seventh Framework Programme FP7/2007-2013/ under REA Grant Agreement No 317089.

\bibliographystyle{JHEP}
\bibliography{bibContact}

\newpage
\section*{Appendix A - Evaluating the free particle determinants}
Here we determine the normalisation constants used in section 2 using $\zeta$-function regularisation. With our gauge choice ($e = \top$) we make a change of variables $t = \top\tau$ and define
\begin{equation}
	\mathcal{Z}^{\prime} = \int_{\boldsymbol{\omega}\left(0\right) = \mathbf{a}}^{\boldsymbol{\omega}\left(T\right) = \mathbf{b}} \mathscr{D}\boldsymbol{\omega} \,e^{-\int_{0}^{T} \frac{\boldsymbol{\dot{\omega}}^{2}}{2} d\tau}
\end{equation}
which can be interpreted as the matrix element $\left<\mathbf{b}\left|e^{-T\mathcal{\hat{H}}}\right|\mathbf{a}\right>$ with Hamiltonian $\mathcal{\hat{H}} = \frac{\hat{p}^{2}}{2}$. In the Sch{\"o}dinger representation this becomes the position space representation of the heat kernel: $\left<\mathbf{b}\left|e^{-T\frac{\boldsymbol{\nabla}^{2}}{2}}\right|\mathbf{a}\right>$. This is a well know expression but we find it using functional methods in keeping with the spirit of the remainder of this article.

Generalising to arbitrary dimension $D$, the integral over ${\omega}$ gives a functional determinant and a boundary term:
\begin{equation}
	\mathcal{Z}^{\prime} =\pi^{\frac{D}{2}} \left( \det{\left( \frac{d^{2}}{dt^{2}}\right)} \right)^{-\frac{D}{2}} e^{-\frac{\left(b - a\right)^{2}}{2T}}.
\end{equation}
The determinant of an operator, $\hat{O}$, can be defined using the $\zeta$-function \cite{Hawk} as
\begin{equation}
	\det{\left(\hat{O}\right)} = \exp{\left(\left.-\frac{d}{dz} \zeta_{\hat{O}}\left(z\right)\right|_{z=0}\right)}\,; \qquad \zeta_{\hat{O}}\left(z\right) \equiv \sum_{n = 1}^{\infty} \lambda_{n}^{-z}
\end{equation}
where the $\lambda_{n}$ are the eigenvalues of $\hat{O}$. The formal product of eigenvalues is then regularised via the analytic continuation of the $\zeta$-function. With Dirichlet boundary conditions the eigenvalues of the operator in $\mathcal{Z}$ are $\lambda_{n} = \left(\frac{n\pi}{T}\right)^{2}$ so
\begin{equation}
	\zeta_{\frac{d^{2}}{dt^{2}}}\left(z\right) = \left(\frac{\pi}{T}\right)^{-2z}\zeta\left(2z\right)
\end{equation}
which has derivative $\zeta^{\prime}\left(0\right) = 2\ln{\left(\frac{T}{\pi}\right)}\zeta\left(0\right) + 2\zeta^{\prime}\left(0\right)$. So with this regularisation we arrive at
\begin{equation}
	\det{\left( \frac{d^{2}}{dt^{2}}\right)} = 2T
\end{equation}
giving
\begin{equation}
	\mathcal{Z}^{\prime} = \left(2\pi T\right)^{-\frac{D}{2}} e^{-\frac{\left(\mathbf{b - a}\right)^{2}}{2T}}.
\end{equation}

If instead we consider mixed boundary conditions $\boldsymbol{\omega}\left(0\right) = 0$, $\dot{\boldsymbol{\omega}}\left(1\right) = 0$ then the eigenfunctions are
\begin{equation}
	\sin{\left(\sqrt{\tilde{\lambda}_{n}}\, t\right)}
\end{equation}
where the eigenvalues are now given by $\tilde{\lambda}_{n} = \left(\frac{\left(2n + 1\right)\pi}{2T}\right)^{2}$. For the $\zeta$-function we now obtain
\begin{equation}
	\tilde{\zeta}_{\frac{d^{2}}{dt^{2}}}\left(z\right) = \left(\frac{\pi}{T}\right)^{-2z}\zeta\left(2z, \frac{1}{2}\right)
\end{equation}
where we've introduced the Hurwitz zeta function $\zeta\left(s, q\right)$. The derivative with respect to $z$ is then $\tilde{\zeta}^{\prime}\left(0\right) = 2\ln{\left(\frac{T}{\pi}\right)}\zeta\left(0, \frac{1}{2}\right) + 2\zeta^{\prime}\left(0, \frac{1}{2}\right)$. Now $\zeta\left(0, \frac{1}{2}\right) = 0$ so the first term vanishes along with the $T$ dependence, leaving
\begin{equation}
	\det{\left( \frac{d^{2}}{dt^{2}}\right)} = 2
\end{equation}
for the case of mixed boundary conditions. The change in boundary conditions also alters the boundary contributions from the classical action and we find that $\mathcal{Z^{\prime}}$ is a constant independent of $T$ and $\mathbf{a}$.

For fermionic fields the kinetic term is first order in derivatives and we impose anti-periodic boundary conditions in the case of closed paths which represent traces. We may of course calculate 
\begin{equation}
	\mathcal{Z}_{F} = \int_{\psi\left(T\right) = -\psi\left(0\right)} \mathscr{D}\psi ~e^{-\int_{0}^{T} \dot{\psi} \cdot \psi d\tau} = \left(\det{\left(\frac{d}{dt}\right)}\right)^{\frac{D}{2}}
\end{equation}
using standard techniques from quantum mechanics but for completeness we continue to apply functional methods. We need the eigenvalues $\lambda_{n} = \frac{\left(2n + 1\right)\pi i}{T}$ and form the determinant via
\begin{equation}
	\prod_{n = -\infty}^{\infty}  \frac{\left(2n + 1\right)\pi i}{T}. 
\end{equation}
The $\zeta$-function for this operator is thus
\begin{equation}
	\left(\frac{2 \pi i}{T}\right)^{-z}\zeta\left(z, \frac{1}{2}\right),
\end{equation}
which has derivative $\zeta^{\prime}\left(0\right) = \ln{\left(\frac{T}{2 \pi i}\right)}\zeta\left(0, \frac{1}{2}\right) + \zeta^{\prime}\left(0, \frac{1}{2}\right)$ which evaluates to $-\ln{\sqrt{2}}$ so
\begin{equation}
	\mathcal{Z}_{F} = 2^{\frac{D}{2}}
\end{equation}
which we note does not depend on $T$ -- indeed the change of variables to $\tau \in \left[0, T\right]$ does not change the form of the kinetic term. 

The other result we need is the normalisation constant for the open path action which includes the term $\int_{0}^{1}d\tau \,\frac{\chi_{0}}{T} \dot{\omega} \cdot \psi$. We use (\ref{gamma}) to calculate
\begin{equation}
	\int d\chi_{0} \delta\left(\chi_{0} - \Xi\right) \mathscr{D}\psi ~e^{-\int_{0}^{T}d\tau \,\frac{1}{2}\dot{\psi} \cdot \psi - \frac{\chi_{0}}{2T}\dot{\omega} \cdot \psi} \propto \int d\chi_{0} \delta\left(\chi_{0} - \Xi\right)\mathscr{T}\left(e^{\int d\tau \frac{\chi_{0}}{2\sqrt{2}T} \dot{\omega} \cdot \gamma}\right).
\end{equation}
Integrating over $\chi_{0}$ picks out
\begin{equation}
	1 + \Xi\frac{\left(b - a\right)\cdot \gamma}{2\sqrt{2}T}
\end{equation}
which is to be multiplied by $\mathcal{Z}_{F}$ calculated above. In the main text we impose $\Xi = 0$ which vastly simplifies the remaining calculations because it decouples $\omega$ and $\psi$. 
\end{document}